\documentclass[11pt,a4paper]{article}
\pdfoutput=1 %

\usepackage{jcappub} %

\usepackage[utf8]{inputenc}
\usepackage[T1]{fontenc}
\usepackage{amsfonts}

\usepackage{microtype}

\usepackage{derivative}
\usepackage[separate-uncertainty=true]{siunitx}

\usepackage[caption=false]{subfig} %

\usepackage{xcolor}

\usepackage{tabularx}

\usepackage{cleveref}

\newcommand*\dd{\mathop{}\!\mathrm{d}}

\newcommand{\LCDM}{$\Lambda$\text{CDM}}
\newcommand{\wowaCDM}{$w_0w_\mathrm{a}$\text{CDM}}
\newcommand{\numass}{\ensuremath{\sum m_\nu}}
\newcommand{\mubest}{\ensuremath{\mu_{\textrm{best}}}}
\newcommand{\muupper}{\ensuremath{\mu_{95}}}
\newcommand{\muupperIH}{\ensuremath{\mu_{95}^{\text{(IO)}}}}
\newcommand{\muupperNH}{\ensuremath{\mu_{95}^{\text{(NO)}}}}
\newcommand{\Lya}{Lyman-$\alpha$}
\makeatletter
\g@addto@macro\bfseries{\boldmath}
\makeatother

\usepackage{orcidlink}

\title{Cosmological neutrino mass: a frequentist overview in light of DESI}

\author[1]{{D.~Chebat}\orcidlink{0009-0006-7300-6616},}
\author[1]{{C.~Yèche}\orcidlink{0000-0001-5146-8533},}
\author[1]{{E.~Armengaud}\orcidlink{0000-0001-7600-5148},}
\author[2,3]{{N.~Sch\"oneberg}\orcidlink{0000-0002-7873-0404},}
\author[2,3]{{M.~Walther}\orcidlink{0000-0002-1748-3745},}
\author[1]{{A.~de~Mattia}\orcidlink{0000-0003-0920-2947},}
\author[4]{{J.~Rohlf}\orcidlink{0000-0001-6423-9799},}
\author[5]{{J.~Aguilar},}
\author[4]{{S.~Ahlen}\orcidlink{0000-0001-6098-7247},}
\author[6,7]{{D.~Bianchi}\orcidlink{0000-0001-9712-0006},}
\author[8]{{D.~Brooks},}
\author[5]{{T.~Claybaugh},}
\author[5]{{A.~Cuceu}\orcidlink{0000-0002-2169-0595},}
\author[9]{{A.~de la Macorra}\orcidlink{0000-0002-1769-1640},}
\author[8]{{P.~Doel},}
\author[5,10]{{S.~Ferraro}\orcidlink{0000-0003-4992-7854},}
\author[11]{{A.~Font-Ribera}\orcidlink{0000-0002-3033-7312},}
\author[12,13]{{J.~E.~Forero-Romero}\orcidlink{0000-0002-2890-3725},}
\author[14,15,16]{{E.~Gaztañaga}\orcidlink{0000-0001-9632-0815},}
\author[17]{{G.~Gutierrez},}
\author[18]{{C.~Hahn}\orcidlink{0000-0003-1197-0902},}
\author[19,1]{{H.~K.~Herrera-Alcantar}\orcidlink{0000-0002-9136-9609},}
\author[20]{{C.~Howlett}\orcidlink{0000-0002-1081-9410},}
\author[21,22]{{D.~Huterer}\orcidlink{0000-0001-6558-0112},}
\author[23]{{M.~Ishak}\orcidlink{0000-0002-6024-466X},}
\author[11]{{J.~Jimenez}\orcidlink{0000-0001-8528-3473},}
\author[24]{{R.~Joyce}\orcidlink{0000-0003-0201-5241},}
\author[24]{{S.~Juneau}\orcidlink{0000-0002-0000-2394},}
\author[25]{{R.~Kehoe},}
\author[26]{{D.~Kirkby}\orcidlink{0000-0002-8828-5463},}
\author[5]{{A.~Kremin}\orcidlink{0000-0001-6356-7424},}
\author[8]{{O.~Lahav},}
\author[5]{{A.~Lambert},}
\author[5]{{M.~Landriau}\orcidlink{0000-0003-1838-8528},}
\author[27]{{L.~Le~Guillou}\orcidlink{0000-0001-7178-8868},}
\author[1]{{C.~Magneville},}
\author[28,11]{{M.~Manera}\orcidlink{0000-0003-4962-8934},}
\author[29,11]{{R.~Miquel},}
\author[30]{{J.~Moustakas}\orcidlink{0000-0002-2733-4559},}
\author[31,32]{{G.~Niz}\orcidlink{0000-0002-1544-8946},}
\author[1,5]{{N.~Palanque-Delabrouille}\orcidlink{0000-0003-3188-784X},}
\author[33,34,35]{{W.~J.~Percival}\orcidlink{0000-0002-0644-5727},}
\author[36]{{F.~Prada}\orcidlink{0000-0001-7145-8674},}
\author[37]{{I.~P\'erez-R\`afols}\orcidlink{0000-0001-6979-0125},}
\author[38]{{G.~Rossi},}
\author[39]{{E.~Sanchez}\orcidlink{0000-0002-9646-8198},}
\author[5]{{D.~Schlegel},}
\author[5]{{J.~Silber}\orcidlink{0000-0002-3461-0320},}
\author[24]{{D.~Sprayberry},}
\author[22]{{G.~Tarl\'{e}}\orcidlink{0000-0003-1704-0781},}
\author[24]{{B.~A.~Weaver},}
\author[27]{{P.~Zarrouk}\orcidlink{0000-0002-7305-9578},}
\author[5]{{R.~Zhou}\orcidlink{0000-0001-5381-4372},}
\author[40]{{H.~Zou}\orcidlink{0000-0002-6684-3997},}

\affiliation[1]{IRFU, CEA, Universit\'{e} Paris-Saclay, F-91191 Gif-sur-Yvette, France}
\affiliation[2]{University Observatory, Faculty of Physics, Ludwig-Maximilians-Universit\"{a}t, Scheinerstr. 1, 81677 M\"{u}nchen, Germany}
\affiliation[3]{Excellence Cluster ORIGINS, Boltzmannstrasse 2, D-85748 Garching, Germany}
\affiliation[4]{Department of Physics, Boston University, 590 Commonwealth Avenue, Boston, MA 02215 USA}
\affiliation[5]{Lawrence Berkeley National Laboratory, 1 Cyclotron Road, Berkeley, CA 94720, USA}
\affiliation[6]{Dipartimento di Fisica ``Aldo Pontremoli'', Universit\`a degli Studi di Milano, Via Celoria 16, I-20133 Milano, Italy}
\affiliation[7]{INAF-Osservatorio Astronomico di Brera, Via Brera 28, 20122 Milano, Italy}
\affiliation[8]{Department of Physics \& Astronomy, University College London, Gower Street, London, WC1E 6BT, UK}
\affiliation[9]{Instituto de F\'{\i}sica, Universidad Nacional Aut\'{o}noma de M\'{e}xico,  Circuito de la Investigaci\'{o}n Cient\'{\i}fica, Ciudad Universitaria, Cd. de M\'{e}xico  C.~P.~04510,  M\'{e}xico}
\affiliation[10]{University of California, Berkeley, 110 Sproul Hall \#5800 Berkeley, CA 94720, USA}
\affiliation[11]{Institut de F\'{i}sica d’Altes Energies (IFAE), The Barcelona Institute of Science and Technology, Edifici Cn, Campus UAB, 08193, Bellaterra (Barcelona), Spain}
\affiliation[12]{Departamento de F\'isica, Universidad de los Andes, Cra. 1 No. 18A-10, Edificio Ip, CP 111711, Bogot\'a, Colombia}
\affiliation[13]{Observatorio Astron\'omico, Universidad de los Andes, Cra. 1 No. 18A-10, Edificio H, CP 111711 Bogot\'a, Colombia}
\affiliation[14]{Institut d'Estudis Espacials de Catalunya (IEEC), c/ Esteve Terradas 1, Edifici RDIT, Campus PMT-UPC, 08860 Castelldefels, Spain}
\affiliation[15]{Institute of Cosmology and Gravitation, University of Portsmouth, Dennis Sciama Building, Portsmouth, PO1 3FX, UK}
\affiliation[16]{Institute of Space Sciences, ICE-CSIC, Campus UAB, Carrer de Can Magrans s/n, 08913 Bellaterra, Barcelona, Spain}
\affiliation[17]{Fermi National Accelerator Laboratory, PO Box 500, Batavia, IL 60510, USA}
\affiliation[18]{Steward Observatory, University of Arizona, 933 N. Cherry Avenue, Tucson, AZ 85721, USA}
\affiliation[19]{Institut d'Astrophysique de Paris. 98 bis boulevard Arago. 75014 Paris, France}
\affiliation[20]{School of Mathematics and Physics, University of Queensland, Brisbane, QLD 4072, Australia}
\affiliation[21]{Department of Physics, University of Michigan, 450 Church Street, Ann Arbor, MI 48109, USA}
\affiliation[22]{University of Michigan, 500 S. State Street, Ann Arbor, MI 48109, USA}
\affiliation[23]{Department of Physics, The University of Texas at Dallas, 800 W. Campbell Rd., Richardson, TX 75080, USA}
\affiliation[24]{NSF NOIRLab, 950 N. Cherry Ave., Tucson, AZ 85719, USA}
\affiliation[25]{Department of Physics, Southern Methodist University, 3215 Daniel Avenue, Dallas, TX 75275, USA}
\affiliation[26]{Department of Physics and Astronomy, University of California, Irvine, 92697, USA}
\affiliation[27]{Sorbonne Universit\'{e}, CNRS/IN2P3, Laboratoire de Physique Nucl\'{e}aire et de Hautes Energies (LPNHE), FR-75005 Paris, France}
\affiliation[28]{Departament de F\'{i}sica, Serra H\'{u}nter, Universitat Aut\`{o}noma de Barcelona, 08193 Bellaterra (Barcelona), Spain}
\affiliation[29]{Instituci\'{o} Catalana de Recerca i Estudis Avan\c{c}ats, Passeig de Llu\'{\i}s Companys, 23, 08010 Barcelona, Spain}
\affiliation[30]{Department of Physics and Astronomy, Siena College, 515 Loudon Road, Loudonville, NY 12211, USA}
\affiliation[31]{Departamento de F\'{\i}sica, DCI-Campus Le\'{o}n, Universidad de Guanajuato, Loma del Bosque 103, Le\'{o}n, Guanajuato C.~P.~37150, M\'{e}xico}
\affiliation[32]{Instituto Avanzado de Cosmolog\'{\i}a A.~C., San Marcos 11 - Atenas 202. Magdalena Contreras. Ciudad de M\'{e}xico C.~P.~10720, M\'{e}xico}
\affiliation[33]{Department of Physics and Astronomy, University of Waterloo, 200 University Ave W, Waterloo, ON N2L 3G1, Canada}
\affiliation[34]{Perimeter Institute for Theoretical Physics, 31 Caroline St. North, Waterloo, ON N2L 2Y5, Canada}
\affiliation[35]{Waterloo Centre for Astrophysics, University of Waterloo, 200 University Ave W, Waterloo, ON N2L 3G1, Canada}
\affiliation[36]{Instituto de Astrof\'{i}sica de Andaluc\'{i}a (CSIC), Glorieta de la Astronom\'{i}a, s/n, E-18008 Granada, Spain}
\affiliation[37]{Departament de F\'isica, EEBE, Universitat Polit\`ecnica de Catalunya, c/Eduard Maristany 10, 08930 Barcelona, Spain}
\affiliation[38]{Department of Physics and Astronomy, Sejong University, 209 Neungdong-ro, Gwangjin-gu, Seoul 05006, Republic of Korea}
\affiliation[39]{CIEMAT, Avenida Complutense 40, E-28040 Madrid, Spain}
\affiliation[40]{National Astronomical Observatories, Chinese Academy of Sciences, A20 Datun Road, Chaoyang District, Beijing, 100101, P.~R.~China}

\emailAdd{domitille.chebat@cea.fr}

\abstract{%
  We derive constraints on the neutrino mass using a variety of recent cosmological datasets, including DESI BAO, the full-shape analysis of the DESI matter power spectrum and the one-dimensional power spectrum of the \Lya{} forest (P1D) from eBOSS quasars
  as well as the cosmic microwave background (CMB).
  The constraints are obtained in the frequentist formalism by constructing profile likelihoods and applying the Feldman-Cousins prescription to compute confidence intervals.
  This method avoids potential prior and volume effects that may arise in a comparable Bayesian analysis.
  Parabolic fits to the profiles allow one to distinguish changes in the upper limits from variations in the constraining power $\sigma$ of the different data combinations. 
  We find that all profiles in the \LCDM{} model are cut off by the $\numass \geq 0$ bound, meaning that the corresponding parabolas reach their minimum in the unphysical sector.
  The most stringent 95\% C.L. upper limit is obtained by the combination of DESI DR2 BAO, \textit{Planck} PR4 and CMB lensing at \qty{53}{\meV}, below the minimum of \qty{59}{\meV} set by the normal ordering.
  The corresponding constraining power $\sigma$ is \qty{43}{\meV}, which highlights the importance of the cut-off by negative values in the determination of the upper limit.
  Extending \LCDM{} to non-zero curvature and \wowaCDM{} relaxes the constraints past \qty{59}{\meV} again, but only \wowaCDM{} exhibits profiles with a minimum at a positive value. 
  Additionally, we extend the formalism to constrain the lightest neutrino mass.
  For DESI DR2 BAO, \textit{Planck} PR4 and CMB lensing, we find confidence limits at \num{20} and \qty{19}{\meV}
  for normal and inverted ordering, respectively.
  Using a combination of DESI DR1 full-shape, BBN and eBOSS \Lya{} P1D, we successfully constrain the neutrino mass independently of the CMB. 
  This combination yields $m_\mathrm{l} \leq \num{97}$ and \qty{98}{\meV} in the normal and inverted orderings, and total neutrino mass $\numass \leq \qty{285}{\meV}$ (95\% C.L.). 
  The addition of DESI full-shape or \Lya{} P1D to CMB and DESI BAO results in small but noticeable improvement of the constraining power of the data.
  \Lya{} free-streaming measurements especially improve the constraint. Since they are based on eBOSS data, this sets a promising precedent for upcoming DESI data.
  }

\begin{document}

\maketitle

\tableofcontents

\section{Introduction}\label{sec:introduction}

Neutrino oscillations were first observed with solar neutrinos~\cite{Davis:1968cp,bahcall-davis,GALLEX:1994rym,SAGE:1994ctc,Super-Kamiokande:1998qwk,SNO:2001kpb,Borexino:2008dzn} and confirmed with atmospheric neutrinos~\cite{super-kamiokandecollaborationEvidenceOscillationAtmospheric1998,AMBROSIO200159,PhysRevD.68.113004,PhysRevD.99.032007,antarescollaboration2019}, accelerators~\cite{PhysRevD.74.072003,PhysRevLett.112.191801,PhysRevLett.125.131802,PhysRevLett.121.139901,C_Rubbia_2011,PhysRevLett.107.041801,PhysRevLett.118.231801}, and reactor experiments~\cite{PhysRevLett.90.021802,PhysRevLett.108.131801,PhysRevLett.108.171803,PhysRevLett.108.191802,2022103927}.
By convention, the neutrino masses are labeled $m_1$, $m_2$ and $m_3$, with $m_1 < m_2$.
The collective experimental results (see Particle Data Group~\cite{PDG2024}) show that at least two of the neutrinos have non-zero masses.
The oscillation experiments are sensitive to the difference between squared masses of the neutrinos, $\Delta m_{21}^2 \sim \qty{7.5 e-5}{\eV^2}$ and $\left|\Delta m_{32}^2\right| \sim \qty{2.5 e-3}{\eV^2}$.
Thus, two of the masses, $m_1$ and $m_2$, are close to each other, while $m_3$ is either larger than $m_1$ and $m_2$ ($m_1 < m_2 < m_3$), referred to as normal ordering (NO), or smaller than $m_1$ and $m_2$ ($m_3 < m_1 < m_2$), referred to as inverted ordering (IO).
By setting the lightest of the three masses to \num{0}, one can determine the smallest possible total mass \numass{} allowed by either ordering,
\begin{equation}\label{eq:mass-minima}
  \left(\numass{}\right)^\mathrm{NO} > \qty{58.980(0.304)e-3}{\eV}, \quad \left(\numass{}\right)^\mathrm{IO} > \qty{99.824(0.581)e-3}{\eV}.
\end{equation}

A cosmic neutrino background with a present-day temperature of \qty{1.9}{K} and number density \qty{340}{cm^{-3}} is a central prediction of the standard model of cosmology (\LCDM{}), and while not directly observed, is essential in the understanding of Big-Bang Nucleosynthesis (BBN), the cosmic microwave background (CMB) anisotropies, and large scale structure in the Universe (see  J. Lesgourgue and L. Verde in section 26 of~\cite{PDG2024}).
Neutrinos have uniquely behaved as radiation at the time of CMB acoustic oscillations and as dark matter at the time of structure formation.
At least two of the three species of cosmic neutrinos are non-relativistic today.
Cosmological data are sensitive to neutrino masses through gravitational effects, providing a remarkable connection between particle physics and cosmology.

Cosmology is primarily sensitive to \numass{}, and largely insensitive to the individual masses themselves~\cite{archidiaconoWhatWillIt2020,lesgourguesMassiveNeutrinosCosmology2006,lesgourguesNeutrinoMassCosmology2012,hannestadNeutrinoPhysicsPrecision2010}, providing complementary information to that from neutrino oscillations.
In the very early universe, neutrinos have large kinetic energies and are highly relativistic.
As time passes, they eventually cool down enough to become non-relativistic, and start contributing to the matter density $\Omega_\mathrm{m}(z)$ rather than to the radiation density $\Omega_\mathrm{r}(z)$.
Neutrinos become non-relativistic when their average momentum approaches their mass.
The redshift  $z_\nu$ at which this transition takes place depends on neutrino masses:
a neutrino of mass $m_\nu$ transitions at 
$z_\nu =m_\nu/(\qty{53e-5}{\eV}) - 1$~\cite{PDG2024}. For $m_\nu = \qty{50}{\meV}$, $z_\nu\sim 100$.
The transition from relativistic to non-relativistic neutrinos, thus, happens in the time span between recombination and large-scale structure formation.
A direct consequence of this transition is the increase of the matter energy density fraction, $\Omega_\mathrm{m}(z)$, compared to the case of massless neutrinos.
This affects the Hubble parameter,
\begin{equation}
  H(z) = H_0 \sqrt{\Omega_\Lambda + \Omega_\mathrm{m} (1+z)^3 + \Omega_\mathrm{r} (1+z)^4} ,
\end{equation}\label{eq:hubble}
where $\Omega_\Lambda$ and $\Omega_\mathrm{r}$ are the dark energy and radiation energy density fractions, respectively,
and is referred to as the geometrical effect.
For a fixed $H_0$, massive neutrinos cause $\Omega_\mathrm{m}(z)$ and $H(z)$ to change relative to a massless neutrino scenario.
In practice, a variation in $H(z)$ can also be attributed to either a change in $H_0$ or in $\Omega_\mathrm{m}$.
Baryon acoustic oscillation (BAO) measurements from the Dark Energy Spectroscopic Instrument (DESI)~\cite{leviDESIExperimentWhitepaper2013,desicollaborationDESIExperimentPart2016,adameDESI2024VI2025,desicollaborationDESIDR2Results2025} constrain $\Omega_\mathrm{m}$ but not $H_0$, and are, thus, unable to constrain the neutrino mass alone.
Similarly, due to the $H_0$ - $\Omega_\mathrm{m}$ degeneracy, CMB measurements such as those from \textit{Planck}~\cite{planckcollaborationPlanck2018Results2020,planckcollaborationPlanck2018Results2020a,choudhuryUpdatedResultsNeutrino2020} or the Atacama Cosmology Telescope (ACT)~\cite{louisAtacamaCosmologyTelescope2025} fail to constrain the geometrical effect.
Since the $H_0$ - $\Omega_\mathrm{m}$ degeneracy can be broken by the DESI constraint on $\Omega_\mathrm{m}$, analyses that combine both datasets become sensitive to the geometrical effect, and thus to the neutrino mass.

Another consequence of non-relativistic neutrinos is that they participate in clustering. 
At the non-relativistic transition, neutrinos are hot dark matter. %
Thus, unlike cold dark matter and baryons, neutrinos are hot enough that they free-stream at small scales and only engage in clustering at scales above some free-streaming scale $\lambda_\mathrm{fs}$.
This results in a suppression of power in the matter power spectrum at scales below $\lambda_\mathrm{fs}$, compared to massless neutrinos~\cite{doroshkevichAstrophysicalImplicationsNeutrino1980,huWeighingNeutrinosGalaxy1998,kiakotouNeutrinoMassDark2008,desicollaborationDESI2024VII2024,ivanovCosmologicalParametersNeutrino2020}.
This effect is commonly referred to as the free-streaming or small-scale suppression effect. 
In current cosmological datasets, it is measurable in two ways.
The reduced clustering weakens the lensing of CMB photons~\cite{lesgourguesProbingNeutrinoMasses2006}, measured by \textit{Planck} and ACT~\cite{madhavacherilAtacamaCosmologyTelescope2024,quAtacamaCosmologyTelescope2024,aghanimPlanck2018Results2020,carronCMBLensingPlanck2022}.
Additionally, improvements in perturbation theory models have allowed exploitation of the full-shape information from large-scale structure (LSS) to constrain the neutrino mass through the measurement of free-streaming~\cite{huWeighingNeutrinosGalaxy1998,desicollaborationDESI2024VII2024,ivanovCosmologicalParametersNeutrino2020}.

In this work, we perform a frequentist analysis based on profile likelihoods~\cite{heroldProfileLikelihoodsCosmology2025} to determine constraints on the neutrino mass from different combinations of cosmological likelihoods.
The profile method enables convenient visual comparison between datasets, and provides information on the statistical strength of the data even in situations where the profile is cut off by the physical $\numass > 0$ bound.
Other works have implemented neutrino mass models that can be effectively extended to negative masses~\cite{craigNoNsGood2024,greenCosmologicalPreferenceNegative2025,elbersNegativeNeutrinoMasses2025,elbersConstraintsNeutrinoPhysics2025} to deal with this limitation.
In our case, although the profiles are graphically extrapolated into the negative sector for ease of visual comparison, computations are strictly kept to $\numass > 0$, and the upper limit computations take this lower bound into account. 
We also derive upper limits constrained by $\qty{0.059}{\eV}$ and $\qty{0.100}{\eV}$ lower bounds, according to the minima imposed by normal and inverted orderings presented in~\cref{eq:mass-minima}.

The rest of our work is organized as follows: in~\cref{sec:methodology}, we describe the frequentist framework used to build and analyze the profile likelihoods and derive the corresponding confidence limits.
\Cref{sec:data-codes} lists the various datasets and codes used for the analysis.
In~\cref{sec:CMB-free-streaming}, we report results based on datasets combinations that strongly rely on geometrical information and where small-scale suppression measurement stems from CMB lensing measurements.
\Cref{sec:FS-free-streaming} focuses on exploiting free-streaming information from full-shape measurements of the matter power spectrum.
Finally, in~\cref{sec:ordering}, we consider individual neutrino masses instead of \numass{}.
We investigate the impact of the mass splitting modeling in the Boltzmann solver when computing profile likelihoods, and consider the impact of shifting minimal \numass{} to that allowed by normal or inverted ordering.
We summarize our findings and conclude in~\cref{sec:conclusion}.

\section{Inference methodology}\label{sec:methodology}

\subsection{Profile likelihoods}\label{subsec:profile-likelihoods}

We use a frequentist approach to report constraints on the sum of neutrino masses \numass{}.
Specifically, we construct profile likelihoods by performing likelihood maximization over a span of \numass{} values.
Let $\mathcal{L}\left(\numass, \mathcal{C}, \mathcal{N}\right)$ be the likelihood of interest where $\mathcal{C}$ is our cosmological basis, excluding \numass{}, and $\mathcal{N}$ are the possible nuisance parameters associated with the likelihood.
The profile likelihood is defined as a function of \numass{}: for a fixed neutrino mass, it is the result of the maximization of the likelihood over all other parameters---cosmological or nuisance.
It is generally considered in log-space, so for a Gaussian likelihood, this corresponds exactly to a $\chi^2$ function, and we refer to the profile likelihoods as $\chi^2(\numass)$.
Since we have a one-dimensional problem, the profiles are simple parabolas.
In the rest of this work, we perform minimization of $-2 \log{\mathcal{L}}$ for several fixed values of \numass{} and fit the resulting points to a parabola.

\begin{figure}[htb!]
  \centering
  \includegraphics[width=\linewidth]{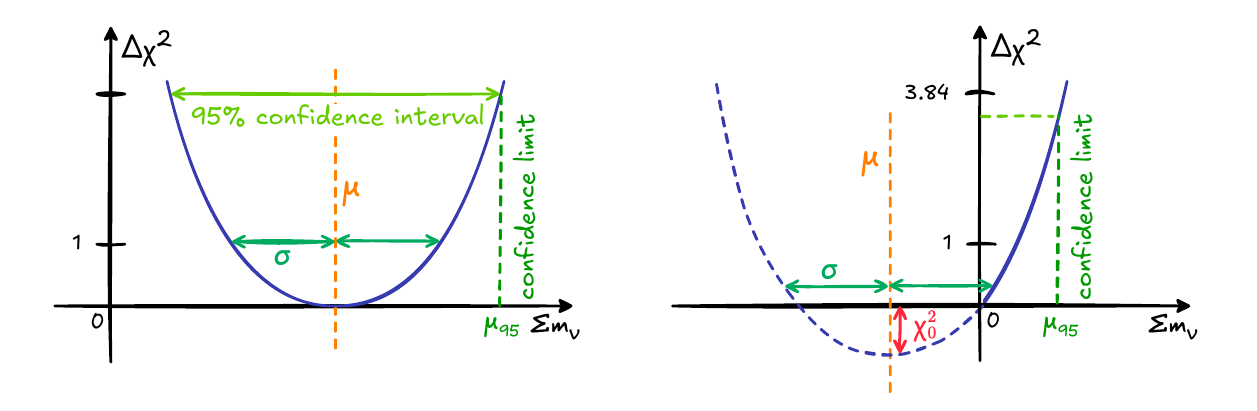}
  \caption{%
    Profile likelihoods in the regular and interrupted case, with parabolic fit parameters.
  }
 \label{fig:PL-drawing}
\end{figure}

The next step is to build a likelihood ratio for the best \numass{}, the neutrino mass that minimizes the profile.
Normally, this amounts to reporting $\Delta\chi^2$ by subtracting the parabola minimum (see~\cref{fig:PL-drawing}).
In the case of the neutrino mass, however, there is a natural lower bound at $\numass = 0$ beyond which we cannot proceed. Furthermore, considering the results of oscillation experiments, one can even argue for a lower bound at \num{59} or \qty{100}{\meV} for the normal and inverted orderings, respectively.
Since cosmological constraints tend to favor very low neutrino masses, most of the profiles we present in~\cref{sec:CMB-free-streaming,sec:FS-free-streaming} are actually cut off by the $\numass = 0$ bound before the minimum of the parabola.
This has been observed in previous works on neutrino mass profile likelihoods~\cite{planckcollaborationPlanckIntermediateResults2014,naredo-tueroCriticalLookCosmological2024}, and is analogous to the Bayesian posteriors
~\cite{elbersConstraintsNeutrinoPhysics2025,desicollaborationDESIDR2Results2025,tristramCosmologicalParametersDerived2024} that do not peak at positive values.
Since the minimum of the parabola is in a forbidden region, we instead report $\Delta\chi^2$ as the difference between $\chi^2$ and the minimal computed $\chi^2$.
In the most frequent case where we can only probe a branch of the parabola, this reduces to
\begin{equation}
  \Delta\chi^2\left(\numass\right) = \chi^2\left(\numass\right) - \chi^2\left(\numass = \qty{0.005}{\eV}\right),
\end{equation}
where the last term has been set to an arbitrary lower mass bound near zero of \qty{0.005}{\eV}.
We fit $\Delta\chi^2$ to a parabola of the form $x \mapsto \left(x - \mu_0\right)^2/\sigma^2 - \chi^2_0$, and report $\mu_0$ and $\sigma$.
The latter can directly be interpreted as the constraining power of the data.
In the case where the parabola is cut off, the offset $\chi^2_0$ corresponds to the tension between $\numass = 0$ and the parabola's minimum.
Graphically, we extend the profile into the $\numass \leq 0$ sector, which allows for easy visual comparison of different profiles.
A simplified example is shown in~\cref{fig:PL-drawing}.

For a standard, uninterrupted profile, 95\% C.L. limits can be reported by finding the mass such that $\Delta\chi^2 = 3.84$.
In the case of cut-off profiles, the overlap between the parabola and negative values make this reasoning invalid.
Instead, we report 95\% C.L. limits following the Feldman-Cousins prescription~\cite{feldmanUnifiedApproachClassical1998} as described in~\cref{subsec:FC-limits}.
Nevertheless, unless $\mu_0$ is negative and several $\sigma$ away from zero, $\Delta\chi^2 = 3.84$ is a reasonable proxy for the 95\% C.L. limit.
We thus display the $\Delta\chi^2 = 3.84$ line on plots to aid with visual comparison.

\subsection{Confidence limits}\label{subsec:FC-limits}
In the case of the neutrino mass, there exists a physical limit $\numass > 0$.
Because current constraints are very close to this boundary, the classical confidence interval often overlaps with negative values, and is thus rendered incorrect.
In~\cite{feldmanUnifiedApproachClassical1998}, the authors describe the construction of a corrected confidence interval for a Gaussian likelihood whose mean should be positive, often called the Feldman-Cousins prescription after the authors.
The prescription can be directly followed to build upper limits for the neutrino mass which take into account that \numass{} must be positive.

Current neutrino mass constraints also overlap with oscillation-based lower limits.
It is thus desirable to build modified upper limits with physical boundaries $\numass \geq \qty{59}{\meV}$ (NO) and $\numass \geq \qty{100}{\meV}$ (IO), which we do by adapting the Feldman-Cousins prescription as described below.

\begin{figure}[htb!]
  \centering
  \includegraphics[width=0.8\linewidth]{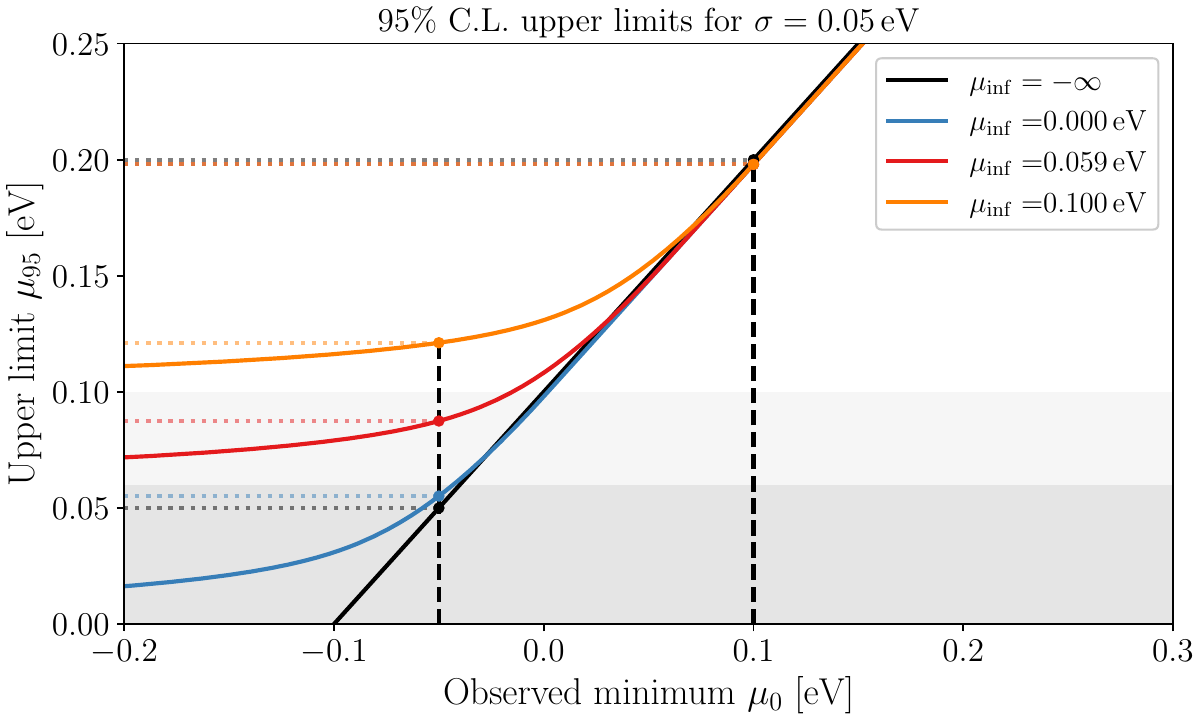}
  \caption{%
    95\% C.L. upper limits at fixed $\sigma = \qty{0.05}{\eV}$, for lower boundaries at \num{0}, \num{0.059} and \qty{0.100}{\eV}, as well as without a lower limit at all.
    The cases $\mu_0 = \qty{-0.05}{\eV}$ and $\mu_0 = \qty{0.10}{\eV}$ are detailed.
    For $\mu_0 \leq \mu_\mathrm{inf} $ (such as $\mu_0 = \qty{-0.05}{\eV}$), the upper limit is increased compared to a boundary-less case.
    This difference progressively vanishes as the observed minimum increases to values above the boundary, and for $\mu_0 = \qty{0.10}{\eV}$, it is barely visible.
  }
 \label{fig:FC-limits}
\end{figure}

Let $\mathcal{P}(x|\mu)$ be a Gaussian likelihood with mean $\mu$ and scale $1$, such that $\mu$ must be greater than some boundary $\mu_\mathrm{inf}$.
We define \mubest{} as the $\mu \geq \mu_\mathrm{inf}$ that maximizes the likelihood for a given $x$, which amounts to $\mubest = \max{(x, \mu_\mathrm{inf})}$.
We then consider the ratio $R(x, \mu)$ of the likelihood $\mathcal{P}(x|\mu)$ to its maximum $\mathcal{P}(x|\mubest)$ as an ordering principle to construct acceptance intervals:
\begin{equation}
  R(x, \mu) = \frac{\mathcal{P}(x|\mu)}{\mathcal{P}(x|\mubest)} = 
  \begin{cases}
    \mathrm{e}^{-\frac{1}{2}(x-\mu)^2} & \text{if $x \geq \mu_\mathrm{inf}$}\\
    \mathrm{e}^{-\frac{1}{2}(x - \mu)^2} / \mathrm{e}^{-\frac{1}{2}(\mu_\mathrm{inf} - \mu)^2} & \text{if $x < \mu_\mathrm{inf}$}
  \end{cases}
\end{equation}
For a coverage probability $\alpha$ and a given $\mu$, the acceptance interval is $[x_1, x_2]$ such that $R(x_1, \mu) = R(x_2, \mu)$ and $\int_{x_1}^{x_2} R(x, \mu) \dd x = \alpha$.
One can then 
determine confidence intervals $[\mu_1, \mu_2]$.
The reasoning can be applied to any Gaussian likelihood of scale $\sigma \neq 1$ by normalizing it and scaling the resulting confidence interval again.
In the following, we will use this construction with $\mu_\mathrm{inf}= 0$, \num{59} and \qty{100}{\meV} to report 95\% confidence intervals named \muupper{}, {\muupperNH{}} and \muupperIH{} respectively.
An example of these different limits is depicted in~\cref{fig:FC-limits} for a given value of $\sigma$ and different profile positions.

\section{Data and software}\label{sec:data-codes}

Cosmological inference is performed in a 7-parameter \LCDM{}+\numass{} model, comprised of $\Omega_\mathrm{m}$, $H_0$, $\ln{\left( 10^{10} A_\mathrm{s} \right)}$, $n_\mathrm{s}$, $\Omega_\mathrm{b} h^2$, $\tau$ and \numass{},
which refer to the matter energy density fraction, the Hubble parameter, the logarithm of the amplitude of primordial density fluctuations, the spectral index, the baryon energy density fraction times $h^2$, where $h = H_0 / \qty{100}{\km.\s^{-1}.\mega pc^{-1}}$, the reionization optical depth, and the summed neutrino masses, respectively.
In the following, we detail data, data products and codes used in our analysis as well as the corresponding nomenclature for the article.

\subsection{Cosmological likelihoods}

\paragraph{Cosmic Microwave Background} 
We make use of cosmic microwave background (CMB) measurements by \textit{Planck}~\cite{planckcollaborationPlanck2018Results2020} and ACT~\cite{louisAtacamaCosmologyTelescope2025}.
We combine \textit{Planck} low-$\ell$ EE and TT likelihoods with high-$\ell$ TT, TE, and EE.
The low-$\ell$ data products are the \texttt{Commander} and \texttt{SimAll} likelihoods from the 2018 \textit{Planck} data release~\cite{planckcollaborationPlanck2018Results2020}.
For high-$\ell$ information, we use both 2018 (PR3) and 2020 (PR4) TTTEEE likelihoods.
In the following, \textit{Planck} PR3 refers to the official \texttt{plik} likelihood from the 2018 data release~\cite{planckcollaborationPlanck2018Results2020}, while \textit{Planck} PR4 corresponds to the \texttt{CamSpec} likelihood based on the \textit{Planck} 2020 analysis~\cite{rosenbergCMBPowerSpectra2022}.

We also use a compressed, foreground-marginalized likelihood for the ACT DR6 high-$\ell$ TTTEEE data~\cite{louisAtacamaCosmologyTelescope2025} referred to as \texttt{ACT-lite}\footnote{\url{https://github.com/ACTCollaboration/DR6-ACT-lite}}.
We follow the ACT collaboration and combine this likelihood with the low-$\ell$ \texttt{Sroll2} \textit{Planck} EE likelihood\footnote{Available at \url{https://sroll20.ias.u-psud.fr/} and \url{https://web.fe.infn.it/~pagano/low_ell_datasets/sroll2/}} described in~\cite{paganoReionizationOpticalDepth2020}.
This combination is referred to as \texttt{ACT-lite} in the following.

Some profiles include isolated \textit{Planck} 2018 information on $n_\mathrm{s}$, $\ln{\left( 10^{10} A_\mathrm{s} \right)}$, or both.
This corresponds to the constraints reported on these parameters in~\cite{planckcollaborationPlanck2018Results2020}, implemented as Gaussian likelihoods.
\begin{itemize}
  \item $n_{\mathrm{s},\ 1\sigma}$ corresponds to the \textit{Planck} constraint.
  \item $n_{\mathrm{s},\ 10\sigma}$ is a relaxed version of the former, with a standard deviation ten times greater.
  \item $n_{\mathrm{s},\ 1\sigma}$ + $A_s$ implements 1-$\sigma$ constraints for both $n_\mathrm{s}$ and $\ln{\left( 10^{10} A_\mathrm{s} \right)}$ as well as the corresponding covariance.
\end{itemize}

\paragraph{CMB lensing}
In addition to TT, TE, EE information on the CMB, we include CMB lensing information both from \textit{Planck} PR4 and ACT DR6~\cite{madhavacherilAtacamaCosmologyTelescope2024,quAtacamaCosmologyTelescope2024} combined in one likelihood\footnote{Specifically, we use the \texttt{actplanck\_baseline} variant in version \texttt{1.2}, available at \url{https://github.com/ACTCollaboration/act_dr6_lenslike}} as described in~\cite{carronCMBLensingPlanck2022}.
CMB lensing is included by default in the profiles labeled \textit{Planck} PR3, \textit{Planck} PR4, and ACT-lite.

\paragraph{Galaxy Clustering} %
We use galaxy clustering measurements by 
DESI to constrain the neutrino mass. 
DESI is a spectroscopic instrument that uses 5000 robotic fiber positioners to capture target spectra over
\qty{7}{\square\deg} field of view~\cite{desicollaborationOverviewInstrumentationDark2022,poppettOverviewFiberSystem2024,desicollaborationDESIExperimentPart2016,millerOpticalCorrectorDark2024,silberRoboticMultiobjectFocal2022}.
Targets are selected from the public Legacy Surveys~\cite{deyOverviewDESILegacy2019,myersTargetselectionPipelineDark2023} and their spectra are classified and processed into a redshift catalog~\cite{guySpectroscopicDataProcessing2023,brodzellerPerformanceQuasarSpectral2023,schlaflySurveyOperationsDark2023}.
After the first round of survey validation for each category of target~\cite{lanDESISurveyValidation2023,alexanderDESISurveyValidation2023,cooperOverviewDESIMilky2023,hahnDESIBrightGalaxy2023,juneauIdentifyingMissingQuasars2025,zhouTargetSelectionValidation2023,raichoorTargetSelectionValidation2023,chaussidonTargetSelectionValidation2023}, cosmological analysis has been performed based on the contents of the first two data releases, the first of which is now publicly available~\cite{desicollaborationDataRelease12025}.

Measurements of BAO help constrain the neutrino mass through geometrical information. 
We employ such measurements from the first two DESI data releases, DR1~\cite{adameDESI2024III2025,adameDESI2024IV2025,adameDESI2024VI2025} and DR2~\cite{desicollaborationDESIDR2Results2025, desicollaborationDESIDR2Results2025a}, and will refer to them as DR1 BAO and DR2 BAO.
Additionally, we make use of the full-shape likelihood of DESI DR1~\cite{desicollaborationDESI2024FullShape2025,desicollaborationDESI2024VII2024}, which is sensitive to neutrino free-streaming, in combination to the associated BAO measurement.
We refer to this combination as DESI FS.

\paragraph{Big Bang Nucleosynthesis}
Baryon abundance in the universe can be determined from BBN. In the following, we use the constraint $\Omega_\mathrm{b}h^2 = \num[group-digits=integer]{0.02218 \pm 0.00055}$ in \LCDM{} from a recent analysis~\cite{schoneberg2024BBNBaryon2024}, and refer to it as BBN.

\paragraph{Supernovae}
We use distance measurements from the sample of type Ia supernovae (SN) published by the \textit{Dark Energy Survey} in their Year 5 data release~\cite{descollaborationDarkEnergySurvey2024}, under the form of the corresponding \texttt{Cobaya} likelihood.
When included, this likelihood is referred to as DES-Y5.
Detailed comparison of the Pantheon+ and Union3 SN data sets has been reported in~\cite{elbersConstraintsNeutrinoPhysics2025} and they are not used in this work.

\paragraph{\texorpdfstring{\Lya{}}{Lyman-⍺} forest}
The one-dimensional power spectrum of the \Lya{} forest (P1D) allows probing small-scale suppression of the matter power spectrum. 
Although DESI has measured the \Lya{} forest P1D~\cite{karacayliDESIDR1Lya2025,ravouxDESIDR1Lya2025}, the interpretation relies on the emulation of sets of high precision hydrodynamical simulations and is still a work in progress.

In the following, we use the results from two sets of such simulations on the parameters $n_\mathrm{Lya}$, $A_\mathrm{Lya}$ of the P1D.
Both works use the quasars from Data Release 14 of the \textit{Sloan Digital Sky Survey}, SDSS DR14~\cite{parisSloanDigitalSky2018}.
The first is described in~\cite{palanque-delabrouilleHintsNeutrinoBounds2020} and referred to as Taylor, while the second results from the work in~\cite{waltherEmulatingLymanAlphaForest2024} and is referred to as Lyssa.
Both likelihoods are implemented as compressed information using the chains from~\cite{waltherEmulatingLymanAlphaForest2024}; for Lyssa, we use the chain that includes a mean transmission prior. 
The results of the minimizations that involve either likelihood lie within the cosmological parameter space covered by the respective simulation grids~\cite{rossiSuiteHydrodynamicalSimulations2014,bordeNewApproachPrecise2014,waltherEmulatingLymanAlphaForest2024}.

\subsection{Inference framework}
All likelihoods are implemented in \texttt{Cobaya}~\cite{torradoCobayaCodeBayesian2021,cobayaASCLsoftware}, using the \texttt{CAMB} Boltzmann solver~\cite{lewisEfficientComputationCosmic2000,howlettCMBPowerSpectrum2012}.
In order to perform the minimization, likelihoods are interfaced with the \texttt{Minuit2}~\cite{jamesMinuitSystemFunction1975} minimizer through its Python frontend \texttt{iminuit}~\cite{dembinskiScikithepIminuit2023}.
The baseline cosmological model is flat \LCDM{} in which neutrinos are modeled as three species of degenerate mass.
The effect of this assumption of degenerate neutrino masses is explored in \cref{sec:ordering}.

The minimization is run in a parameter space consisting of the cosmological basis ($\Omega_\mathrm{m}$, $H_0$, $\ln{\left( 10^{10} A_\mathrm{s} \right)}$, $n_\mathrm{s}$, $\Omega_\mathrm{b} h^2$, $\tau$, \numass{}) as well as any nuisance parameters required by the likelihoods involved.
In~\cref{subsec:CMB-w0wa}, we consider extensions to the base \LCDM{}+\numass{} model:
\begin{itemize}
  \item The Chevallier-Polarski-Linder (\wowaCDM{}) dynamical dark energy~\cite{chevallierACCELERATINGUNIVERSESSCALING2001,linderExploringExpansionHistory2003}, in which the equation of state of dark energy is parametrized as $ w(z) = w_0 + \left(z/(1+z) \right) w_\mathrm{a}$.
  \item A universe with non-zero curvature, where $\Omega_\mathrm{K}$ is free.
\end{itemize}
In those cases, we extend the basis to include $w_0$ and $w_\mathrm{a}$ or $\Omega_\mathrm{K}$.
Depending on the choice of likelihoods, the dimension of the parameter space for the minimization may range from 8 to more than 40.
The cosmological basis has been chosen for easy compatibility with the likelihoods used in this work.
It should however be noted that unlike Bayesian inference techniques, such as Markov chain Monte Carlo, minimization --- and, thus, profile likelihoods --- are invariant under a reparametrization of the parameter basis.

It can happen that the whole cosmological basis is not needed. For instance, in the data combinations of~\cref{sec:FS-free-streaming}, the function to minimize is completely independent of $\tau$. In such situations, we freeze the parameter in question.
The CMB and DESI full-shape likelihoods introduce numerous nuisance parameters.
For these, we implement penalties and limits on the explored parameter space as prescribed in their original frameworks.
The cosmological parameters are allowed to vary within the domain described in~\cref{tab:param-domain}.
The nominal observed values of the parameters are summarized by  O. Lahav and A. R. Liddle in section 25 of~\cite{PDG2024}.

\begin{table}[htb!]
  \centering
  \begin{tabular}{lccccccccc}
    \hline
            & $\Omega_\mathrm{m}$ & $H_0$ & $\log{A}$ & $n_\mathrm{s}$ & $\Omega_\mathrm{b} h^2$ & $\tau$ & $w_0$ & $w_\mathrm{a}$ & $\Omega_\mathrm{K}$ \\
    \hline
    Minimum & 0.2                 & 60    & 2.8       & 0.92           & 0.015                   & 0.02   & -3.0  & -3.0           & -0.3                \\
    Maximum & 0.4                 & 90    & 3.9       & 1.02           & 0.03                    & 0.1    & 0.0   & 2.0            & 0.3                 \\
    \hline
\end{tabular}

  \caption{%
    Minimization domain for each cosmological parameter.
  }
\label{tab:param-domain}
\end{table}

\section{Geometry and small-scale suppression with the CMB}\label{sec:CMB-free-streaming}

In this section, we present profiles in the \LCDM{} and \wowaCDM{} frameworks using combinations of BAO and CMB data.
The constraints derive from the geometrical effect as well as measurement  of small-scale suppression from the CMB.

\subsection{Flat \texorpdfstring{\LCDM{}}{ΛCDM}}\label{subsec:CMB-LCDM}

\subsubsection{CMB and DESI BAO}\label{subsubsec:CMB-DESI-LCDM}

\begin{figure}[htb!]
  \centering
  \includegraphics[width=0.8\linewidth]{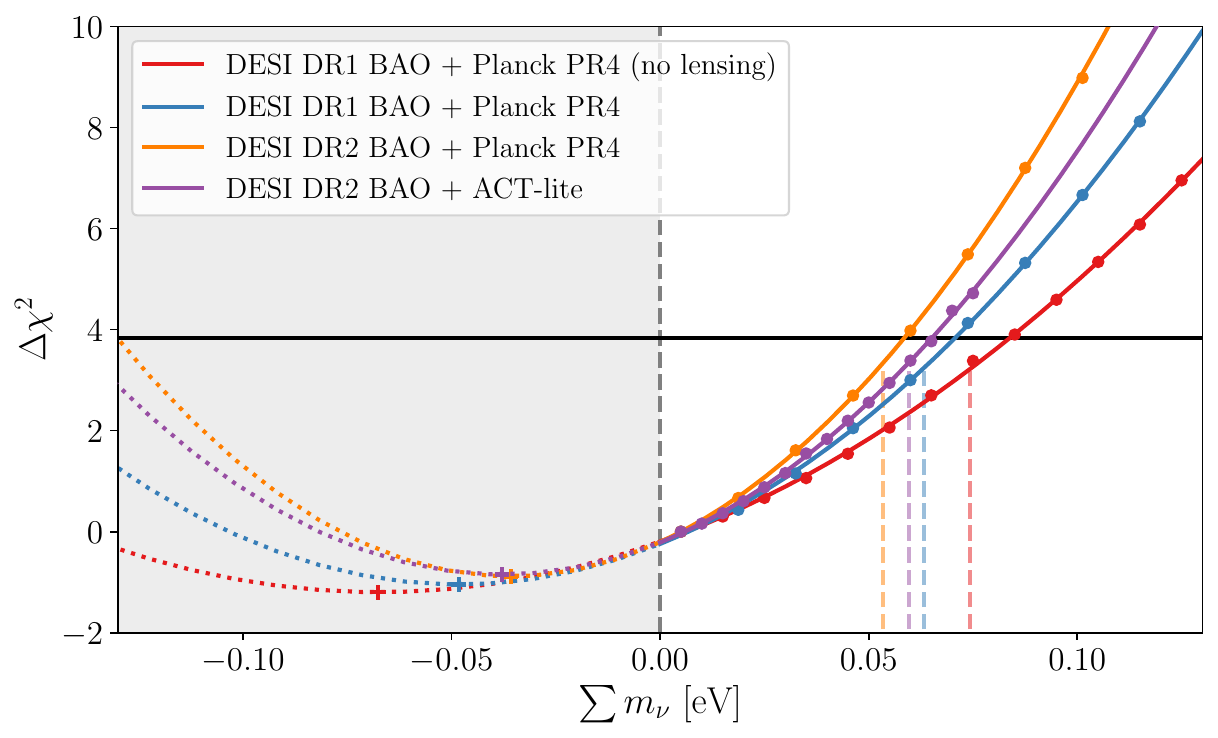}
  \caption{%
    Profile likelihoods for different combinations of BAO with CMB data from \textit{Planck} and ACT.
    Including lensing information and switching from DESI DR1 BAO to DESI DR2 BAO improves the constraining power of the data, and shifts the minima toward the positive values.
    Combined, these effects lead to a lower upper limit.
    Replacing CMB data from \textit{Planck} with that from ACT-lite slightly relaxes the constraint from DESI DR2 BAO + \textit{Planck} PR4 without shifting the parabola.
  }
  \label{fig:DESI-BAO-CMB-LCDM}
\end{figure}

As explained in~\cref{sec:introduction}, combining BAO and CMB data breaks parameter degeneracies specific to each dataset and allows one to constrain the neutrino mass through the geometrical effect.
Including CMB lensing information adds sensitivity to the neutrino mass through a different effect, the suppression of small scales in the matter power spectrum.
Recent BAO data are available from DR1 and DR2 of DESI, and CMB data from \textit{Planck} and ACT.
In~\cref{fig:DESI-BAO-CMB-LCDM}, we compare various combinations of these datasets.
Several conclusions can be drawn from one-on-one comparisons of the profiles.

First, in the most constraining combination (DESI DR2 BAO + \textit{Planck} PR4), replacing \textit{Planck} PR4 with ACT-lite yields little difference.
The upper limit is slightly increased, which is consistent with results reported by Bayesian analyses~\cite{calabreseAtacamaCosmologyTelescope2025,garcia-quinteroCosmologicalImplicationsDESI2025}.
The increase is only due to a slight relaxation of the parabola, whose position does not change.
Since a significant part of the ACT likelihood is shared with Planck, including CMB lensing, this is not unexpected.\footnote{CMB lensing is a combination of \textit{Planck} and ACT data, see \cref{sec:data-codes}. The lower-$\ell$ part of the high-$\ell$ ACT likelihood is taken from \textit{Planck} PR3~\cite{louisAtacamaCosmologyTelescope2025}.}

On the contrary, the switch from DR1 to DR2 BAO and the addition of CMB lensing each cause the parabola minimum to shift, toward the positive.
At constant $\sigma$, this should cause the upper limit to increase.
However, the gain in constraining power offered by these changes is large enough that the upper limit actually decreases in each case.
The addition of CMB lensing allows sensitivity to the free-streaming effect, bringing $\sigma$ down by \qty{13}{\meV}.
The upgrade from DR1 to DR2 for the BAO measurement tightens the constraint on $\Omega_\mathrm{m}$, leading to an \qty{11}{\meV} decrease of $\sigma$.
In both cases, the improvement in constraining power is higher than the change in upper limit could have suggested.

In the end, the most constraining combination (DESI DR2 BAO + \textit{Planck} PR4) yields $\sigma = \qty{43}{\meV}$ and a 95\% C.L. limit at $\numass < \qty{53}{\meV}$.
To put things in perspective, the final design report of DESI originally forecast DESI Y5 to achieve $\sigma \sim \qtyrange[range-units=single]{20}{30}{\meV}$~\cite{desicollaborationDESIExperimentPart2016a}.
The tension with regard to positive values, indicated by the vertical offset of the parabola $\chi^2_0$, remains below \num{1} in all cases and tends to decrease with the inclusion of new data.

\begin{figure}[htb!]
  \centering
  \includegraphics[width=0.8\linewidth]{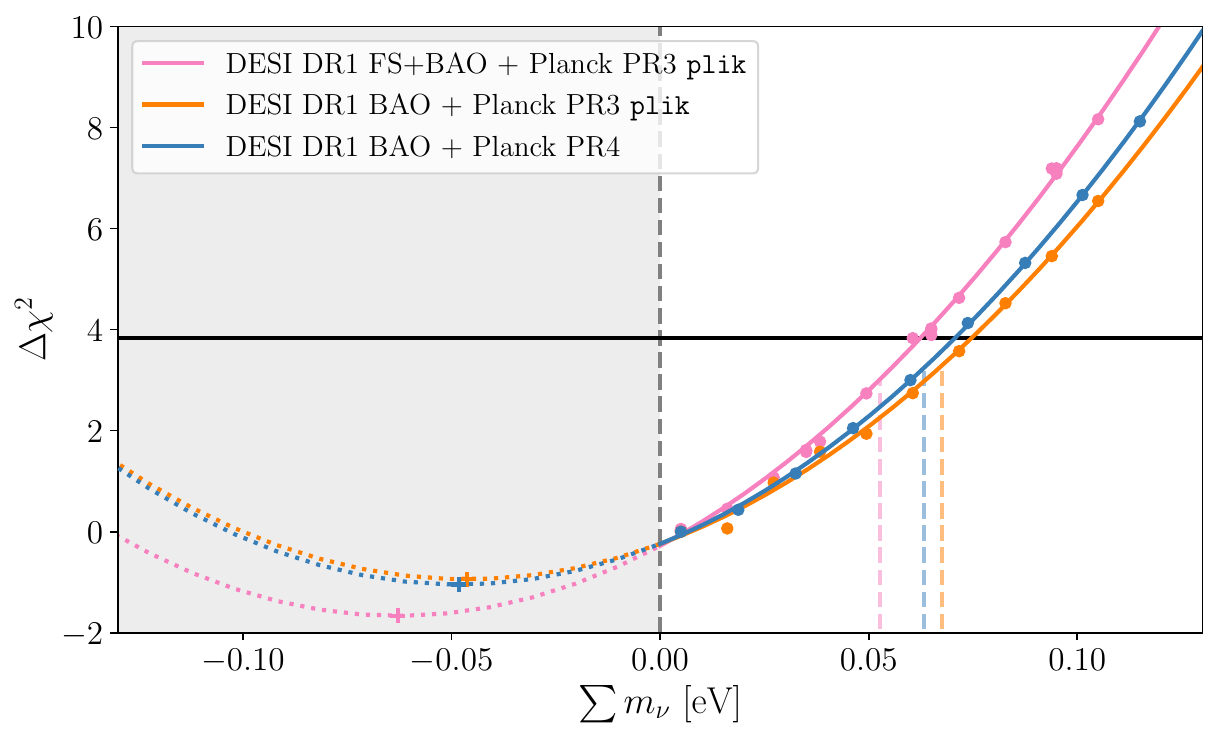}
  \caption{%
    Profile likelihoods for different combinations of DESI BAO and full-shape with CMB data from Planck.
    The change in \textit{Planck} likelihood, from PR3 \texttt{plik} to PR4 \texttt{CamSpec}, produces extremely similar results.
    Adding full-shape information to DR1 BAO shifts the parabola slightly to the negative, with some improvement to the constraining power.
  }
  \label{fig:DESI-FS-CMB-LCDM}
\end{figure}

In addition to the BAO measurement, we can also use the full-shape information from the DESI data. 
Like CMB lensing, the full-shape measurement of the power spectrum should contribute to the constraint through the free-streaming effect.
We follow the Bayesian analysis in the original DESI publication~\cite{desicollaborationDESI2024VII2024} and consider the combination of DESI full-shape and BAO from DR1 with \textit{Planck} PR3 \texttt{plik}.
The corresponding profiles are plotted in~\cref{fig:DESI-FS-CMB-LCDM}.
We first note that the change from \textit{Planck} PR3 \texttt{plik} to \textit{Planck} PR4 \texttt{CamSpec} has very little impact on the profile, save for a slight decrease of the upper limit, also observed in the corresponding Bayesian analyses with DESI DR2~\cite{elbersConstraintsNeutrinoPhysics2025}.

The addition of full-shape information causes the profile to shift toward the negative by \qty{17}{\meV}, while the constraining power barely improves by \qty{2}{\meV}.
As a consequence, the upper limit is pulled down from \num{68} to \qty{53}{\meV}, just like the DESI DR2 BAO + \textit{Planck} PR4 combination.
However, unlike DESI DR2 BAO, the gain from DESI DR1 BAO to full-shape is predominantly driven by the shift of the parabola.
At $\sigma=\qty{53}{\meV}$, the data is about as constraining with full-shape as it is without.
This indicates that small-scale suppression measurements from DESI full-shape are not competitive with CMB lensing yet.
Lastly, in the case of the DESI full-shape profile, there is more significant tension between the parabola's minimum and positive values.

\subsubsection{CMB and \texorpdfstring{\Lya{}}{Lyman-⍺}}\label{subsubsec:CMB-Lya-LCDM}

\begin{figure}[htb!]
  \centering
  \includegraphics[width=0.8\linewidth]{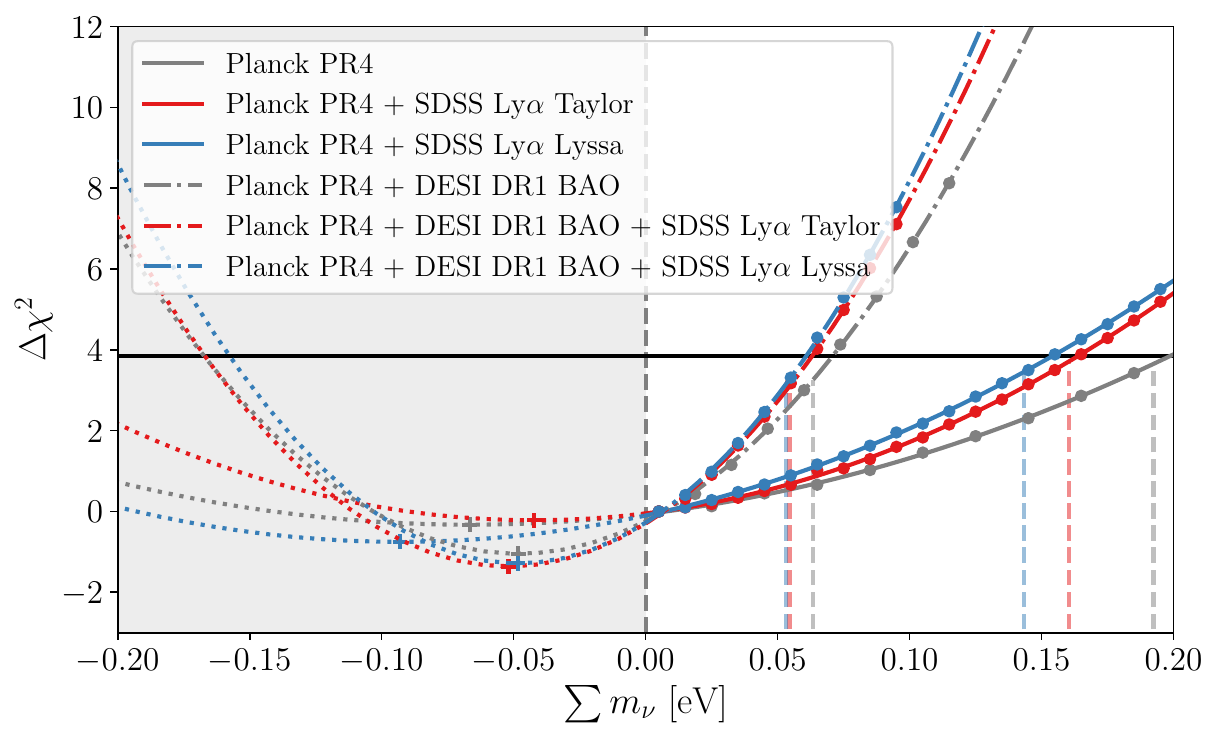}
  \caption{%
    Profile likelihoods for combinations of CMB information from \textit{Planck} PR4, BAO from DESI DR1 and \Lya{} likelihoods ~\cite{palanque-delabrouilleHintsNeutrinoBounds2020,waltherEmulatingLymanAlphaForest2024}.
    Both \Lya{} likelihoods produce similar results, especially when combined with DESI BAO.
    The addition of \Lya{} consistently improves the constraining power.
  }
  \label{fig:CMB-Lya}
\end{figure}

Profiles for the combination of CMB, BAO, and \Lya{} P1D are shown in~\cref{fig:CMB-Lya}.
The addition of either Lyssa or Taylor information to CMB alone results in a decreased upper limit and a smaller $\sigma$.
Although the combination with Taylor results in a higher limit (\qty{160}{\meV}) than Lyssa (\qty{143}{\meV}), it actually carries the tightest constraining power, standing at $\sigma = \qty{102}{\meV}$ compared to the \qty{115}{\meV} of Lyssa.
This stems from the different directions of the shift of the minima compared to the CMB alone: positive for Taylor and negative for Lyssa.
Constraints on $n_\mathrm{Lya}$ and $A_\mathrm{Lya}$ are of the same order or weaker than the ones obtained from \textit{Planck}, 
so while it was not likely that introducing the \Lya{} information would constitute a huge improvement, nevertheless a shift in the limit was observed.

Although both \Lya{} likelihoods are based on the same eBOSS data, as noted in~\cite{waltherEmulatingLymanAlphaForest2024}, the differences in simulations and methodology lead to different results.
Typically, the authors find Lyssa $n_\mathrm{Lya}$ and Taylor $A_\mathrm{Lya}$ agree well with \textit{Planck}, while the Lyssa $A_\mathrm{Lya}$ and Taylor $n_\mathrm{Lya}$ show some tension with \textit{Planck}.
The optimization performed for CMB + \Lya{} profiles show a similar trend.

We also combine \textit{Planck} PR4 and \Lya{} likelihoods with DESI DR1 BAO, to compare the profiles to the combination of \textit{Planck} PR4, DESI DR1 BAO and DESI DR1 full-shape from~\cref{subsubsec:CMB-DESI-LCDM}.
As for DESI full-shape, the upper limit is slightly pulled down by the addition of free-streaming information to the BAO+CMB combination, going from \qty{63}{\meV} to \num{55} and \qty{53}{\meV} for Taylor and Lyssa, respectively.
However, unlike the case of the full-shape, for which the change in upper limit stemmed almost entirely from a shift of the parabola, the addition of \Lya{} introduces minimal shifts of the minimum.
The change is driven by the tightening of $\sigma$, which decreases from \qty{54}{\meV} to \num{50} (Taylor) and \qty{48}{\meV} (Lyssa).
Although the improvement remains small compared to the constraining power without the \Lya{} information, one can expect significant improvement from the upcoming P1D analysis of DESI data.

\subsection{Extensions beyond flat \texorpdfstring{\LCDM{}}{ΛCDM}}\label{subsec:CMB-w0wa}

As previous analyses have noted, neutrino mass constraints exhibit an important model dependence~\cite{adePlanck2013Results2014}.
In the following, we explore two extensions to the flat \LCDM{} model that was used in~\cref{subsec:CMB-LCDM}: dynamical dark energy, and non-zero curvature.

Although the introduction of a cosmological constant $\Lambda$ continues to be the standard for cosmological analysis, recent data have sparked strong interest in the possibility of dynamical dark energy~\cite{desicollaborationDESIDR2Results2025,descollaborationDarkEnergySurvey2025,rubinUnionUNITYCosmology2024,lodhaExtendedDarkEnergy2025}.
It is commonly modeled by introducing the CPL parametrization of the equation of state of dark energy (\cref{sec:data-codes})
and the modified model is referred to as \wowaCDM{}.
The combination of the most recent observations of the BAO feature, the CMB and Type Ia supernovae show that tensions between datasets are alleviated in \wowaCDM{}, leading to a preference over \LCDM{}~\cite{desicollaborationDESIDR2Results2025}.
The same data, when used for Bayesian inference in the determination of the neutrino mass, yield one-dimensional posteriors that peak at positive values.

\begin{figure}[htb!]
  \centering
  \includegraphics[width=0.8\linewidth]{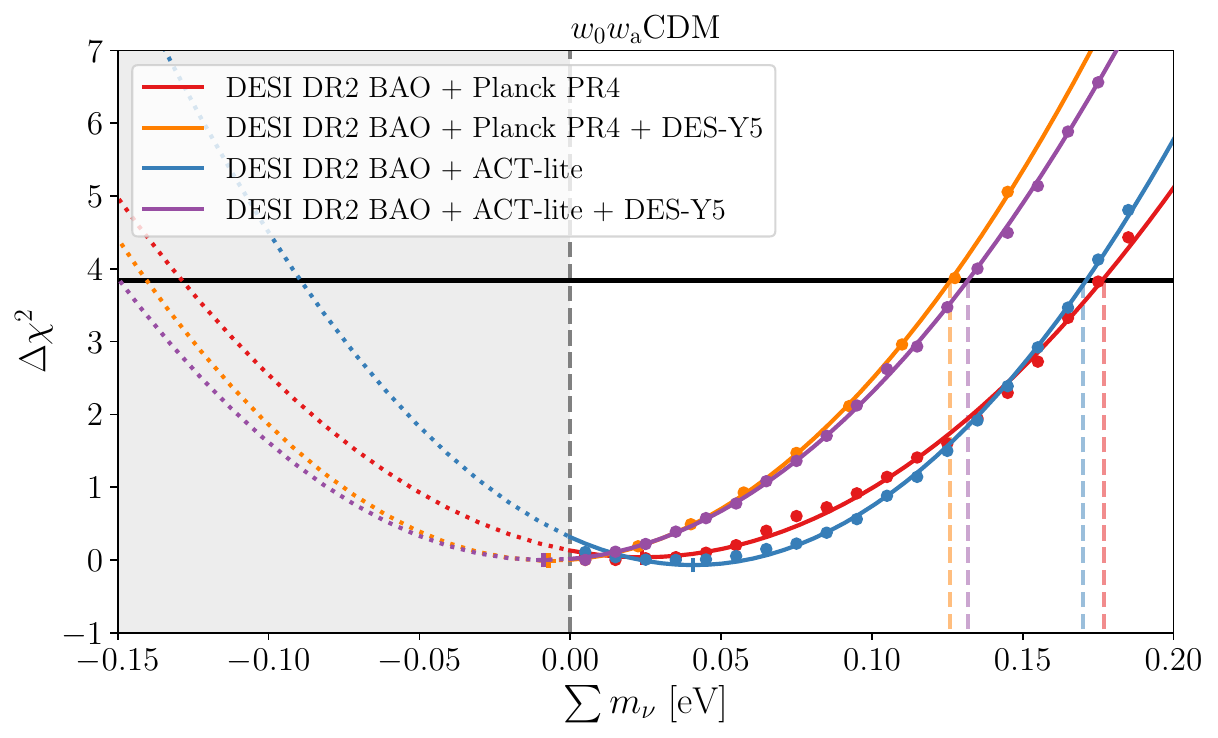}
  \caption{%
    Profile likelihoods for combinations of BAO data, CMB data from \textit{Planck} PR4 and SN from DES-Y5, in a \wowaCDM{} cosmological model.
    DESI DR2 BAO + CMB profiles now exhibit a minimum at a positive value.
    Adding SN information stabilizes both profiles in the same location, but the minimum moves back to a negative value. 
  }
  \label{fig:DESI-CMB-w0waCDM}
\end{figure}

Here, we investigate the effect of dynamical dark energy on the more stringent DESI DR2 BAO + CMB combinations of the previous section.
Since \wowaCDM{} introduces new degeneracies in the analysis, we expect the resulting constraint to be weakened.
The introduction of data from Type Ia SN partially mitigates this loss of constraining power. 
\Cref{fig:DESI-CMB-w0waCDM} presents the corresponding profile likelihoods.
The most glaring difference with \LCDM{} is that both DESI DR2 BAO + \textit{Planck} PR4 and DESI DR2 BAO + ACT-lite now exhibit a minimum at positive values, at \num{24} and \qty{41}{\meV} respectively.
This effect was also observed in \cite{elbersConstraintsNeutrinoPhysics2025}.
As a consequence, the upper limit from the Feldman-Cousins prescription now coincides with $\Delta\chi^2 = 3.84$.
Opening up the dark energy model to \wowaCDM{} largely relaxes the constraints compared to \LCDM{}. 
Although the parabolic fit is worse in \wowaCDM{}, we note that $\sigma$ goes from \num{43} to about \qty{78}{\meV} for the DESI DR2 BAO + \textit{Planck} PR4 combination.
This broadening, as well as the shift of the minima, both contribute to the relaxation of the upper limit to $\numass \leq \qty{177}{\meV}$.
The inclusion of SN data from DES-Y5 pushes the minima slightly to a negative value and stabilizes both curves in the same location.
The ACT-lite combination shows slightly looser constraining power than the \textit{Planck} PR4 one.
This situation is fully analogous to that presented in \LCDM{} (\cref{subsubsec:CMB-DESI-LCDM}).

\begin{figure}[htb!]
  \centering
  \includegraphics[width=0.8\linewidth]{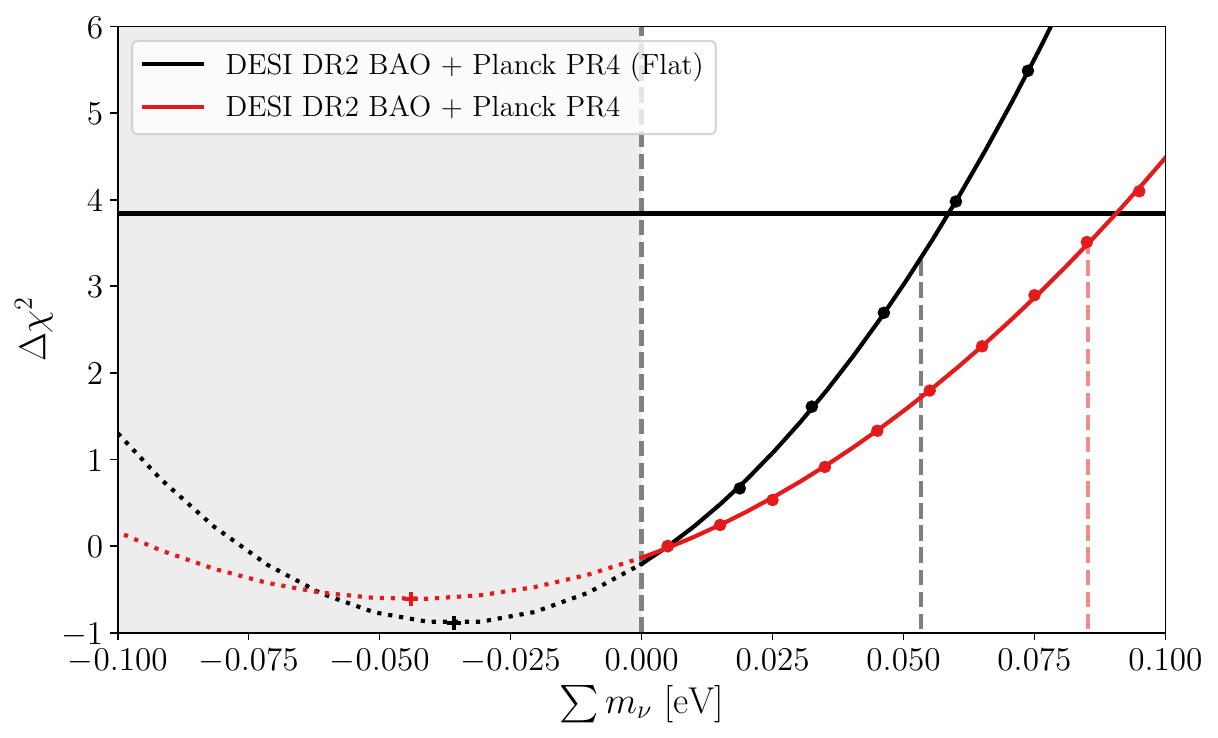}
  \caption{%
    Profile likelihoods for DESI DR2 BAO + \textit{Planck} PR4 in flat and non-flat \LCDM{} cosmological models.
  }
  \label{fig:curvature}
\end{figure}

\paragraph{}
Our base \LCDM{} model assumes flat spacetime.
This hypothesis is commonly assumed in modern cosmological analyses and is supported by observation~\cite{adameDESI2024VI2025}.
Simple inflation models lead to zero curvature, and measurements of $\Omega_\mathrm{K}$ by the CMB alone or combined with BAO are consistent with a flat universe~\cite{desicollaborationDESIDR2Results2025,planckcollaborationPlanck2018Results2020a,efstathiouEvidenceSpatiallyFlat2020,rosenbergCMBPowerSpectra2022}.
These measurements are generally carried out at fixed neutrino mass; however, $\Omega_\mathrm{K}$ is degenerate with $\numass{}$, in particular through the geometrical effect used to determine the neutrino mass.
Analyses conducted with free curvature, thus, find relaxed constraints for BAO and CMB combinations. In~\cite{chenItsAllOk2025}, the authors perform Bayesian inference and the recovered limits become compatible with the inverted ordering minimal mass again.
In~\cref{fig:curvature}, we allow non-zero curvature by letting the parameter $\Omega_\mathrm{K}$ be free.

As $\Omega_\mathrm{K}$ can play a similar role to that of the neutrino mass in the geometrical effect, allowing non-zero curvature should decrease the constraining power of the data. 
Indeed, for the combination of DESI DR2 BAO and \textit{Planck} PR4, varying the curvature degrades $\sigma$ from \qty{43}{\meV} to \qty{64}{\meV}.
The upper limit increases to \qty{85}{\meV}, which is high enough to permit normal hierarchy,
however, the minimum moves further into the negative.

Indeed, DESI DR2 BAO + \textit{Planck} PR4 favors positive $\Omega_\mathrm{K}$ and only marginally higher $\Omega_\mathrm{m}$ than flat \LCDM{}~\cite{desicollaborationDESIDR2Results2025}. 
As a consequence, it also should prefer even lower neutrino mass to compensate for the increase in $\Omega_\mathrm{K}$.
This preference does not appear in the upper limit determination due to the degeneracy on the geometrical effect weakening the constraint, but it is visible in the overall shift of the profile.

The fit neutrino parameters are summarized in \cref{tab:small-scale-lensing}.

\begin{table}[htb!]
  \centering
  \begin{tabularx}{0.9\textwidth}{Xccc}
    \hline
                                               & $\mu_0$ & $\sigma$ & $\muupper$ \\
    \hline
    \hline
    \LCDM{}                                                                      \\
    \hline
    DESI DR1 BAO + Planck PR4 (no lensing)     & -0.068  & 0.067    & 0.074      \\
    DESI DR1 BAO + Planck PR4                  & -0.048  & 0.054    & 0.063      \\
    DESI DR2 BAO + Planck PR4                  & -0.036  & 0.043    & 0.053      \\
    DESI DR2 BAO + ACT-lite                    & -0.038  & 0.048    & 0.060      \\
    \hline
    DESI DR1 BAO + Planck PR3 \texttt{plik}    & -0.046  & 0.055    & 0.068      \\
    DESI DR1 FS+BAO + Planck PR3 \texttt{plik} & -0.063  & 0.053    & 0.053      \\
    \hline
    Planck PR4                                 & -0.067  & 0.130    & 0.192      \\
    Planck PR4 + SDSS \Lya{} Taylor                        & -0.042  & 0.102    & 0.160      \\
    Planck PR4 + SDSS \Lya{} Lyssa                         & -0.093  & 0.115    & 0.143      \\
    Planck PR4 + SDSS \Lya{} Taylor + DESI DR1 BAO         & -0.052  & 0.050    & 0.055      \\
    Planck PR4 + SDSS \Lya{} Lyssa  + DESI DR1 BAO         & -0.048  & 0.048    & 0.053      \\
    \hline
    \hline
    \wowaCDM{}                                                                   \\
    \hline
    DESI DR2 BAO + Planck PR4                  & 0.024   & 0.078    & 0.177      \\
    DESI DR2 BAO + Planck PR4 + DES-Y5         & -0.007  & 0.068    & 0.126      \\
    DESI DR2 BAO + ACT-lite                    & 0.041   & 0.066    & 0.171      \\
    DESI DR2 BAO + ACT-lite + DES-Y5           & -0.009  & 0.077    & 0.132      \\
    \hline
    \hline
    $\Omega_\mathrm{K}$ + \LCDM{}                                                \\
    \hline
    DESI DR2 BAO + Planck PR4                  & -0.044  & 0.064    & 0.085      \\
    \hline
\end{tabularx}

  \caption{%
    Parabolic fits and derived parameters for data combinations in~\cref{sec:CMB-free-streaming}.
    Parameters are reported in \unit{\eV}.
    Unless specified otherwise, CMB data includes CMB lensing.
  }
\label{tab:small-scale-lensing}
\end{table}

\section{Small-scale suppression with large-scale structures}\label{sec:FS-free-streaming}

In~\cref{sec:CMB-free-streaming}, the constraint on neutrino mass is driven by a combination of geometrical information from the DESI BAO data and the \textit{Planck} CMB observations, as well as measurements of the impact of neutrino free-streaming through CMB lensing measurements by \textit{Planck} and ACT.
When including full-shape information from DESI on top of BAO information, there is no significant improvement, which suggests that free-streaming constraints from CMB lensing information dominate the free-streaming constraints from large-scale structure measurements.
In this section, we aim to derive constraints using as little free-streaming information from CMB as possible.

Since the high-$\ell$ polarization and temperature spectra of the CMB are still sensitive to the lensing effect, we do not use the full CMB information and compress the data to the slope and amplitude of the primordial power spectrum.
We start from the DESI DR1 full-shape + BAO likelihood, and add information on $\Omega_\mathrm{b}h^2$ from BBN as well as the CMB slope and amplitude description.
The latter can be obtained from the \textit{Planck} dataset by introducing its constraints on $n_\mathrm{s}$ and $A_\mathrm{s}$, or by considering the quantities $n_\mathrm{Lya}$ and $A_\mathrm{Lya}$ from the simulation grids of the \Lya{} forest 1D power spectrum.
In the \LCDM{} framework, these approaches should be equivalent.

\subsection{Constraints on \texorpdfstring{\numass{}}{∑m\_ν} with compressed primordial information from the CMB}\label{subsec:FS-CMB}

\begin{figure}[htb!]
  \centering
  \includegraphics[width=0.8\linewidth]{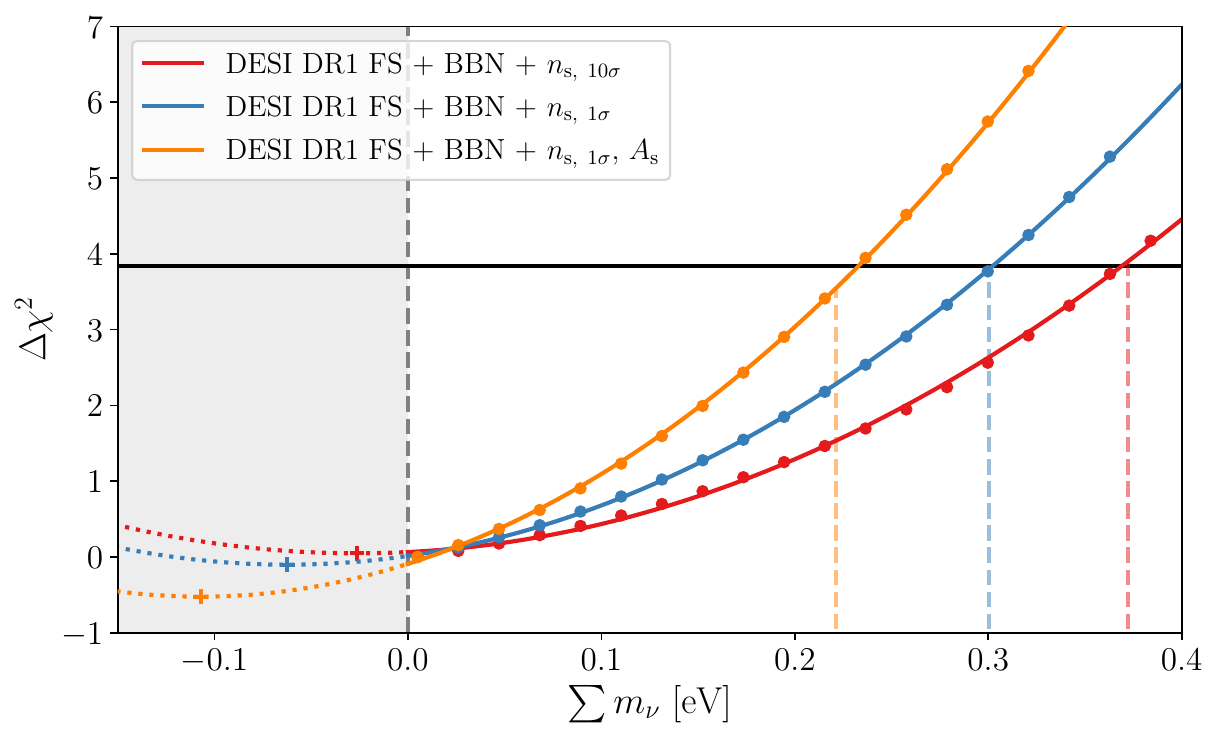}
  \caption{Combination of DESI full-shape (FS), BBN, and \textit{Planck} information on $n_\mathrm{s}$ and $A_\mathrm{s}$.
  Adding more constraining information from \textit{Planck} drives a more stringent upper limit, but also tips the curves to negative values.
  The width of the parabolas $\sigma$ also steadily decreases, which also contributes to the lowering of the upper limit.
  }
  \label{fig:FS-BBN-CMB}
\end{figure}

\Cref{fig:FS-BBN-CMB} presents the profiles for combinations of DESI full-shape and \textit{Planck} information on the primordial power spectrum.
We expect the switch from a loosened constraint on $n_\mathrm{s}$ to the original \textit{Planck} one, and then the inclusion of $A_\mathrm{s}$ to improve the constraining power of the data combination.

The fit parameters reported in~\cref{tab:small-scale-FS} indeed show that $\sigma$ improves by about \qty{20}{\meV} each time, going from \num{203} to \num{184} and then \qty{163}{\meV}.
We also observe a strong decrease of the upper limit, by about \qty{75}{\meV}.
The tightening of the parabola alone only contributes to about half of this shift; the rest is caused by the displacement of the parabola minimum toward the negative.
Although important, especially after the inclusion of $A_\mathrm{s}$, this shift leaves the profiles in reasonable tension with $\numass \geq 0$, at $\chi_0^2 \sim 0.5$ at most.

Compared to the \LCDM{} constraints given by the combination of BAO and CMB data, both constraining power and upper limits are greatly relaxed here.
Several factors are at play in this relaxation, and can be broken down in two categories depending on the neutrino mass effect---geometrical or free-streaming---that they affect.

First and foremost, the data combination does not include the CMB lensing measurement of small-scale suppression, which we determined in~\cref{subsubsec:CMB-DESI-LCDM} to be much more constraining than DESI full-shape.
In the case in where the CMB lensing likelihood was not included, the constraining power in~\cref{subsubsec:CMB-DESI-LCDM} was about three times stronger.
This is partly because, as mentioned above, even without including the lensing in the likelihood, the CMB spectra are sensitive to the lensing at small angular scales and, thus, to the free-streaming effect.
Planck PR4 without the lensing is able to set a 68\% C.L. upper limit at $\numass < \qty{0.161}{\eV}$~\cite{rosenbergCMBPowerSpectra2022}, which is of the same order as the 68\% C.L. confidence limit of the loosest profile in~\cref{fig:FS-BBN-CMB}.
When reducing the information to $n_\mathrm{s}$ and $A_\mathrm{s}$, this sensitivity is severely limited.

The remaining difference lies in disappearance of the geometrical effect.
Since we do not include the full CMB data in this section, there is essentially no geometrical constraint that comes into play, as highlighted in fig.~1 of~\cite{elbersConstraintsNeutrinoPhysics2025}.

\subsection{Constraints on \texorpdfstring{\numass{}}{∑m\_ν} with compressed primordial information from the \texorpdfstring{\Lya{}}{Lyman-⍺} forest}\label{FS-Lya}

\begin{figure}[htb!]
  \centering
  \includegraphics[width=0.8\linewidth]{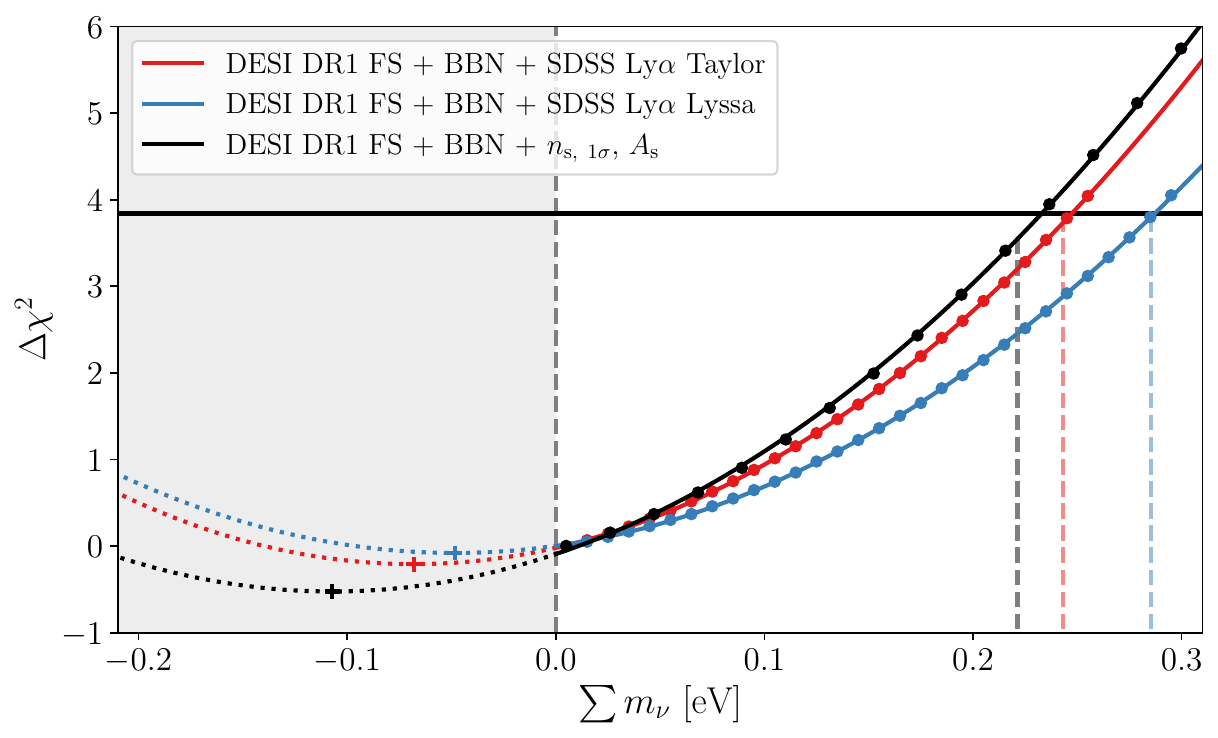}
  \caption{%
    Combination of DESI full-shape (FS), BBN, and \Lya{} likelihoods from either the Taylor (red) or Lyssa (blue) simulation grids.
    All combinations result in comparable minimum positions. 
    The Taylor-based profile is more constraining than the Lyssa one.
    The combination with \textit{Planck} $n_\mathrm{s}$ and $A_\mathrm{s}$ from~\cref{subsec:FS-CMB} (black) is included for comparison.
  }
  \label{fig:FS-BBN-Lya}
\end{figure}

In~\cref{fig:FS-BBN-Lya}, we present profiles for combinations of DESI full-shape with BBN and several \Lya{} likelihoods.
The corresponding fit parameters are reported in~\cref{tab:small-scale-FS}.
The cosmological parameter basis used in the Taylor and Lyssa analyses~\cite{palanque-delabrouilleHintsNeutrinoBounds2020,waltherEmulatingLymanAlphaForest2024} is composed of $\Omega_\mathrm{m}$ and $H_0$ in addition to $n_\mathrm{Lya}$ and $A_\mathrm{Lya}$.
The \Lya{} likelihoods do not provide significant constraints on $\Omega_\mathrm{m}$ and $H_0$, especially compared to the other likelihoods included in the analysis.
In addition, both of these parameters are almost uncorrelated to $n_\mathrm{Lya}$ and $A_\mathrm{Lya}$\footnote{For Lyssa: Around 3\% correlation at most, except for $\Omega_\mathrm{m}$ and $n_\mathrm{Lya}$ at about 20\%.}.
For these reasons, we remove them from the analysis, and expect this should not strongly impact the constraint.

The Taylor profile shown in~\cref{fig:FS-BBN-Lya} is more constraining than its Lyssa counterpart. 
Since both likelihoods are based on the data release 14 of SDSS, this is only attributable to a difference in modeling; either to the inclusion of a prior on $H_0$ in Taylor, or to the emulation differences discussed in the Lyssa publication~\cite{waltherEmulatingLymanAlphaForest2024}.

The constraining powers $\sigma$ of the combinations with either $n_\mathrm{s}$ and $A_\mathrm{s}$ from \textit{Planck}, \qty{163}{\meV}, or $n_\mathrm{Lya}$ and $A_\mathrm{Lya}$ from the \Lya{} forest are comparable: \qty{169}{\meV} and \qty{157}{\meV} with the Lyssa and Taylor likelihoods, respectively. 
Although the profile based on \Lya{} provides a less stringent upper limit, this is almost entirely attributable to a shift in the minimum position and not to a difference in statistical strength.
Indeed, the \textit{Planck}-based profile is shifted to the negative compared to \Lya{}.
We expect that this shift is mostly driven by the difference in the spectral index constrained by each dataset; both \Lya{} likelihood combinations are minimized by lower values of $n_\mathrm{s}$ than the one constrained by \textit{Planck}, which can be compensated by a higher neutrino mass.

\begin{table}[htb!]
  \centering
  \begin{tabularx}{0.9\textwidth}{Xccc}
    \hline
                                                        & $\mu_0$ & $\sigma$ & $\muupper$ \\
    \hline
    DESI FS + BBN + $n_{\mathrm{s},\ 10\sigma}$         & -0.026  & 0.203    & 0.372      \\
    DESI FS + BBN + $n_{\mathrm{s},\ 1\sigma}$          & -0.062  & 0.184    & 0.300      \\
    DESI FS + BBN + {$n_{\mathrm{s},\ 1\sigma}$, $A_s$} & -0.107  & 0.163    & 0.221      \\
    \hline
    DESI FS + BBN + SDSS \Lya{} Taylor                              & -0.068  & 0.157    & 0.243      \\
    DESI FS + BBN + SDSS \Lya{} Lyssa                               & -0.048  & 0.169    & 0.285      \\
    \hline
\end{tabularx}

  \caption{%
    Parabolic fit parameters for data combinations in~\cref{sec:FS-free-streaming}.
    Parameters are reported in \unit{\eV}.
  }
  \label{tab:small-scale-FS}
\end{table}

\section{Individual neutrino masses}\label{sec:ordering}

\subsection{Mass splitting model}

All profile likelihoods presented in previous sections are computed under the so-called degenerate mass assumption, in which there are three massive neutrinos such that
\begin{equation}
  m_1 = m_2 = m_3 = \frac{1}{3} \numass.
\end{equation}
Although we know from oscillation experiments that this hypothesis is not true, it is a standard way of modeling the mass splitting. 
It has been shown to be a reasonable approximation for cosmological determination of the neutrino mass~\cite{couchotCosmologicalConstraintsNeutrino2017,archidiaconoWhatWillIt2020,heroldRevisitingImpactNeutrino2025}, and is the one assumed in recent major data releases and their neutrino analysis, although they also considered mass splittings~\cite{elbersConstraintsNeutrinoPhysics2025,desicollaborationDESIDR2Results2025,calabreseAtacamaCosmologyTelescope2025}.
Other notable approaches include the one originally taken in the \textit{Planck} data releases, which report results in \LCDM{} with a fixed total mass $\numass = \qty{0.059}{\eV}$ made up of one massive neutrino only, and set the two other particles to be massless. 
This approach continues to be taken in current major data releases such as~\cite{planckcollaborationPlanck2018Results2020a,desicollaborationDESIDR2Results2025,louisAtacamaCosmologyTelescope2025}.

Compared to the mass splitting dictated by the normal or inverted ordering, the degenerate approximation breaks down when the total mass approaches the boundaries set by either scheme.
In this region, the lightest neutrino mass $m_\mathrm{l}$ becomes much smaller than the heaviest one,
however, cosmological data is primarily influenced by the total mass \numass{}, which does not depend on the modeling.
In this section, we endeavor to study how big of an impact the modeling choice can have on cosmological analyses.

\begin{figure}[htb!]
  \centering
  \includegraphics[width=0.6\linewidth]{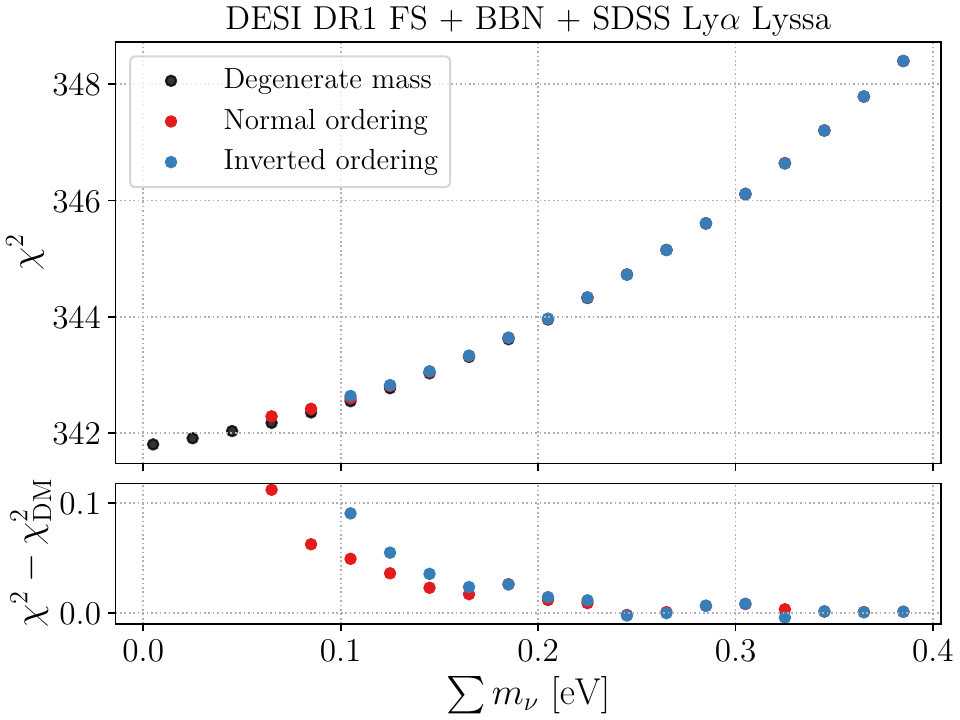}
  \caption{%
   Non-normalized profile likelihoods for DESI full-shape, BBN and information on $n_\mathrm{Lya}$, $A_\mathrm{Lya}$ from the Lyssa likelihood, for different mass splitting modelings in \texttt{CAMB}.
  }
  \label{fig:DM-NH-IH}
\end{figure}

As briefly discussed in~\cref{sec:introduction}, cosmology is almost insensitive to the individual neutrino masses.
They mostly affect the non-relativistic transition redshift $z_\nu$, since heavier neutrinos transition earlier.
Keeping in mind that $z_\nu = m_\nu / (\qty{53e-5}{\eV}) - 1$~\cite{PDG2024}, we are assured that at least two of the three species have completed their non-relativistic transition between recombination and recent times, and are participating in the gravitational effects described in~\cref{sec:introduction}.
Heavier neutrinos can start affecting clustering and contributing to $\Omega_\mathrm{m}(z)$ earlier, however, which modifies the small-scale suppression effect.
The free-streaming scale $\lambda_\mathrm{fs}$ also technically depends on the individual neutrino masses, further affecting the small scales of the matter power spectrum.
In order to gauge whether the degenerate mass assumption is still valid, we compute a profile likelihood for a given set of information and varied the mass splitting model in the Boltzmann solver, \texttt{CAMB}.
The results are presented in~\cref{fig:DM-NH-IH}.

As expected, the biggest variation between modelings happens very close to the respective boundaries of the orderings.
There, the difference in $\chi^2$ compared to a degenerate mass modeling reaches 0.11. %
Generally speaking, the degenerate mass assumption leads to a slight overestimation of the maximized likelihood close to the ordering boundaries.
As a consequence, the minima of the profiles shift towards positive values, and the parabolas close up slightly. 
The effect on $\sigma$ is limited, however, with variations of 8\% at most, as visible in~\cref{tab:CAMB-modeling} where we report parabola parameters and the associated upper limits.
Similar orders of variation were observed for other likelihood combinations of comparable constraining power, such as the combination of DESI full-shape and \textit{Planck} measurements of the primordial power spectrum.
For such data combinations, the degenerate mass modeling exhibits reasonable differences with the inverted and normal modelings.
The effect on the conclusions of the analysis, especially the upper limit $\muupperIH$ or $\muupperNH$, remain constrained to a few percents.

On the other hand, for more constraining data combinations, such a test is not suitable anymore, for several reasons.
Indeed, profiles such as those featured in~\cref{sec:CMB-free-streaming} result in 95\% C.L. limits below the inverted ordering limit, and even below the normal ordering limit in some cases.
Consequently, there is no doubt that comparable results cannot be obtained with the inverted or normal modeling.
Additionally, it is not conceivable to produce a profile from such constraining data combinations, all the while requiring $\numass \geq \qty{59}{\meV}$ or $\qty{100}{\meV}$.
Since such masses are very distant from the parabola's minimum, the analysis would be limited to a region where the parabolic fit cannot be expected to hold up.

\begin{table}[htb!]
  \centering
  \begin{tabularx}{\textwidth}{X|ccc|ccc|ccc}
  \hline
             & $\mu_0$ & $\sigma$ & \muupper & $\mu_0$ & $\sigma$ & \muupperNH & $\mu_0$ & $\sigma$ & \muupperIH \\
  \hline
  Degenerate & -0.049  & 0.170    & 0.285      & -0.042  & 0.167    & 0.294                    & -0.039  & 0.166    & 0.302                    \\
  Normal     & -       & -        & -          & -0.019  & 0.160    & 0.299                    & -0.022  & 0.161    & 0.306                    \\
  Inverse    & -       & -        & -          & -       & -        & -                        & -0.010  & 0.157    & 0.309                    \\
  \hline
\end{tabularx}

  \caption{%
    Parabolic fit parameters for DESI full-shape, BBN and information on $n_\mathrm{Lya}$, $A_\mathrm{Lya}$ from the Lyssa likelihood with different mass splitting models in \texttt{CAMB}, as plotted in~\cref{fig:DM-NH-IH}.
    Parameters are reported in \unit{\eV}.
  }
\label{tab:CAMB-modeling}
\end{table}

\subsection{Lightest neutrino mass}

\begin{figure}[htb!]
  \centering
  \subfloat[DESI full-shape, BBN, \Lya{}]{%
    \includegraphics[width=0.49\linewidth]{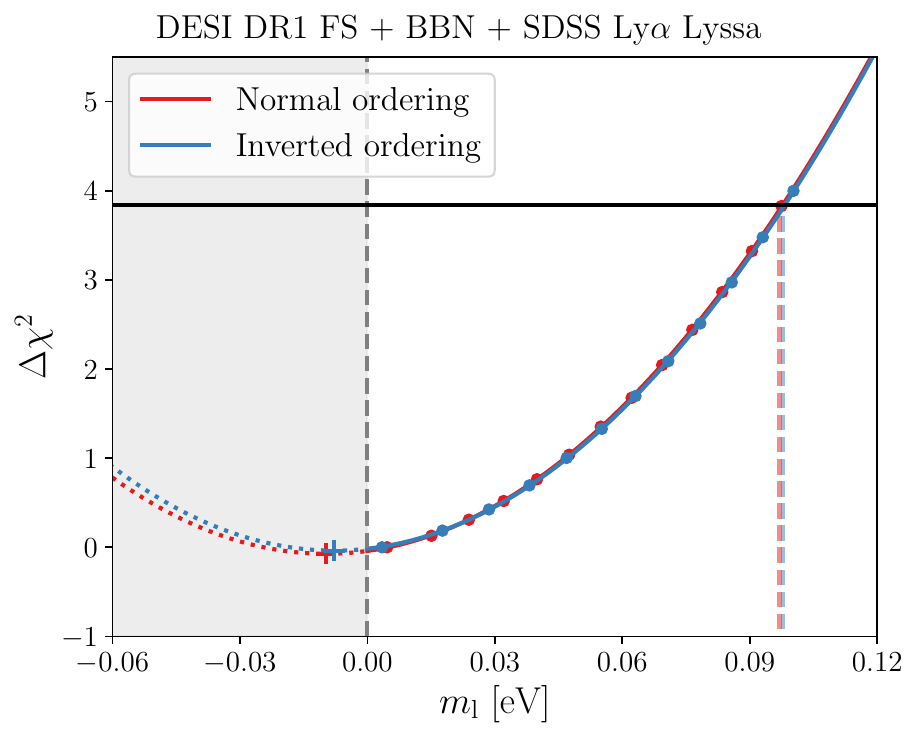}
  }
  \subfloat[DESI DR2 BAO, \textit{Planck} PR4]{%
  \includegraphics[width=0.49\linewidth]{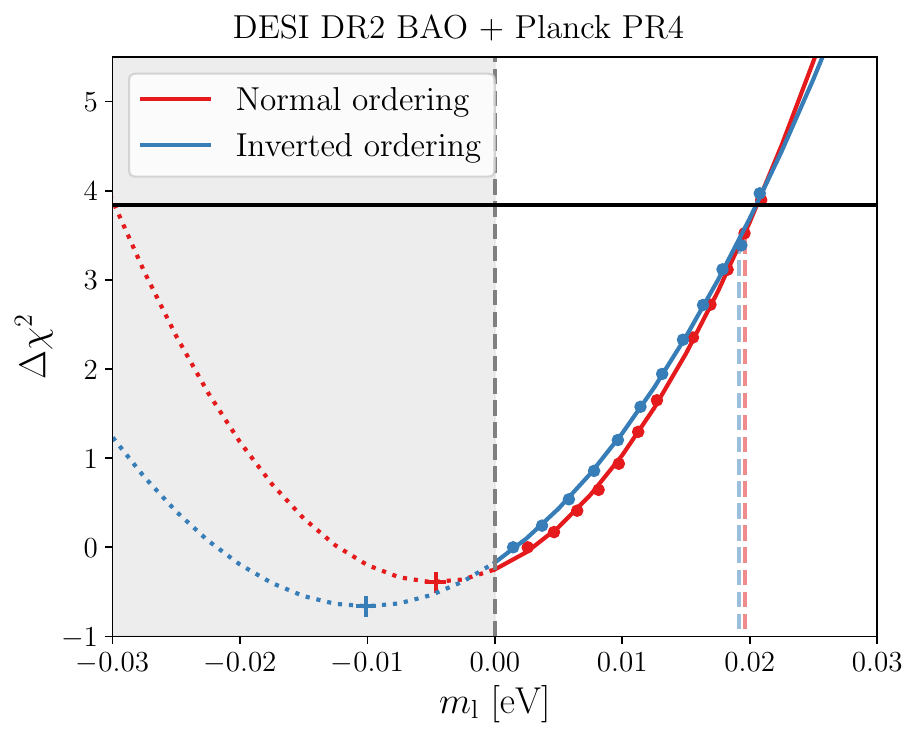}
  }
  \caption{%
    Profile likelihoods for the lightest neutrino mass $m_\mathrm{l}$ based on two likelihood combinations: either DESI full-shape, BBN and Lyssa, or the most constraining DESI DR2 BAO and \textit{Planck} PR4.
  }
  \label{fig:lightest-mass}
\end{figure}

When working with neutrino mass in cosmology, \numass{} is a natural parameter choice as it is more directly constrained.
However, for a given modeling of the mass splitting, which informs us on mass differences, the constraint on \numass{} may be equivalently translated to one on the mass of the lightest neutrino.
For the normal ordering, the lightest neutrino has mass $m_1$, while in the inverted ordering, its mass is $m_3$.
We refer to the lightest mass indifferently as $m_\mathrm{l}$.
Translated results for two likelihood combinations are presented on \cref{fig:lightest-mass}.

In the case of the more relaxed combination which involves DESI full-shape, BBN and Lyssa \Lya{}, in both normal and inverted ordering, the minimum of the parabola is at a negative value, and the constraints are very close.
This behavior is consistent with the previous \numass{} profiles, which preferred a total mass as small as possible. %
The similarity of the profiles is owed to the similarity of the normal and inverted ordering profiles in~\cref{fig:DM-NH-IH}.
The ranges $m_\mathrm{l} \in [0, \muupper{}]$ correspond to $\numass \in [\qty{0.059}{\eV}, \qty{0.305}{\eV}]$ and $\numass \in [\qty{0.100}{\eV}, \qty{0.320}{\eV}]$, so there is important overlap in a region where the $\chi^2$ values are very similar.
The 95\% upper limits derived from the profiles, for DESI full-shape, BBN and the \Lya{} Lyssa likelihood on $n_\mathrm{Lya}$ and $A_\mathrm{Lya}$ are
\begin{align}
  \begin{split}
  & m_\mathrm{l} < \qty{0.097}{\eV} \quad \text{(95\%, Normal ordering)} \\
  & m_\mathrm{l} < \qty{0.098}{\eV} \quad \text{(95\%, Inverted ordering).}
  \end{split}
\end{align}

When applying the same process to the most constraining combination of DESI DR2 BAO and \textit{Planck} PR4, although the upper limits are once again very similar, the profiles themselves differ.
Indeed, due to the tightness of the constraint, $m_\mathrm{l}$ evolves in a domain where the corresponding \numass{} ranges in normal and inverted ordering do not overlap: approximately $[\qty{0.059}{\eV}, \qty{0.095}{\eV}]$ and $[\qty{0.100}{\eV}, \qty{0.130}{\eV}]$.
The minima are once again at negative values, consistently with the preference for masses below the ordering minima.
The 95\% upper limits from DESI DR2 BAO and \textit{Planck} PR4 are
\begin{align}
  \begin{split}
  & m_\mathrm{l} < \qty{0.020}{\eV} \quad \text{(95\%, Normal ordering)} \\
  & m_\mathrm{l} < \qty{0.019}{\eV} \quad \text{(95\%, Inverted ordering).}
  \end{split}
\end{align}
These results are in agreement with the similar analysis carried out in the Bayesian formalism in~\cite{elbersConstraintsNeutrinoPhysics2025}, which found 95\% credible limits at \num{23} and \qty{24}{\meV} respectively for the same data combination.

\begin{table}[htb!]
  \centering
  \begin{tabular}{lcccc}
    \hline
                     & $\mu_0$ & $\sigma$ & $\muupper$ \\
    \hline
    DESI full-shape + BBN + Lyssa $n_\mathrm{Lya}$, $A_\mathrm{Lya}$ \\
    \hline
    Normal ordering  & -0.009  & 0.054    & 0.097      \\
    Inverted ordering & -0.007  & 0.053    & 0.098      \\
    \hline
    DESI DR2 BAO + Planck PR4 \\
    \hline
    Normal ordering & -0.005 & 0.012 & 0.020 \\ 
    Inverted ordering & -0.010 & 0.014 & 0.019 \\
    \hline
  \end{tabular}
  \caption{%
    Parabolic fit parameters for the lightest neutrino mass profiles of~\cref{fig:lightest-mass}.
    Parameters are reported in \unit{\eV}.
  }
\label{tab:lightest-mass}
\end{table}

\section{Conclusion}\label{sec:conclusion}

With recent cosmological data releases, the upper bound on the sum of neutrino masses \numass{} is steadily decreasing.
Combination of major probes such as BAO, CMB, and CMB lensing now produce credible intervals that exclude the minimal mass imposed by the inverted ordering, and continue to creep closer to the normal hierarchy bound of \qty{59}{\meV}.
The interpretation of these upper bounds is complicated by the proximity to the physical bound at $\numass = 0$.
Furthermore, the development of EFT-based likelihoods such as DESI full-shape and the rising interest in alternatives to \LCDM{} has caused the investigated parameter spaces to expand greatly, leading to legitimate concern that volume effects may affect Bayesian analyses.

In this work, we have approached the problem from a frequentist point of view, and chosen to construct profile likelihoods for the neutrino mass.
Such an approach is devoid of any prior and volume effects, and independent of the choice of cosmological basis.
The closeness of the zero mass bound can be accounted for in the confidence limit by applying the Feldman-Cousins prescription~\cite{feldmanUnifiedApproachClassical1998}.
Additionally, profiles of Gaussian likelihood take the shape of a parabola, whose parameters can be fit and used to determine the constraining power $\sigma$ of the profiled data separately from the confidence limit.
This constitutes a powerful tool in the comparison of different datasets.

We find that in a base \LCDM{}+\numass{} cosmological model, using a variety of dataset combinations, all profiles are cut off by the zero mass bound, and the corresponding parabolas reach their minima at unphysical mass ($\numass < 0$).

We first investigate classic data combinations of BAO and CMB data, wherein the bulk of the constraint comes from the geometrical effect of the neutrino mass, with a free-streaming contribution from CMB lensing.
The most stringent data combination, made up of DESI DR2 BAO, CMB from \textit{Planck} PR4, and CMB lensing from ACT DR6 and \textit{Planck} PR4, yields a 95\% C.L. limit of \qty{53}{\meV}, which falls below both normal and inverted ordering minima.
Changes in the CMB likelihood, such as considering \textit{Planck} PR3 or ACT DR6, leave the profiles mostly unaffected aside from a slight loosening.
By contrast, the addition of free-streaming measurements by the DESI DR1 full-shape analysis highlights the distinction between the confidence limit and the constraining power $\sigma$.
Indeed, the confidence limit is artificially pulled down by the shift of the whole parabola, even though the constraining power shows limited gain.
Free-streaming information can also be found in the \Lya{} P1D.
Actually, despite currently stemming from SDSS data, the \Lya{} likelihood already shows interesting added value on $\sigma$, comparable to or even exceeding that of DESI full-shape.

The determination of the neutrino mass through cosmological means is inherently model dependent.
We consider two extensions to the \LCDM{} model: non-zero spatial curvature and dynamical dark energy. 
These extensions have long been studied and have recently seen renewed interest, especially in the context of the neutrino mass~\cite{chenItsAllOk2025,elbersConstraintsNeutrinoPhysics2025}.
Overall, constraints in such extensions are very relaxed by new degeneracies, and all confidence limits now allow at least the normal ordering. 
Only \wowaCDM{} shows a migration of the profiles in the direction of physical mass ($\numass \geq 0$), even exhibiting a trough firmly at positive values in some cases.

Additionally, we considered novel data combinations to derive neutrino mass constraints that rely on free-streaming measurement by large scale structure probes exclusively. 
This is made possible by combining the DESI full-shape and BAO information to BBN data and \Lya{} P1D measurements.
We find that we are able to constrain the neutrino mass independently of the CMB to $\numass \leq \qty{0.243}{\eV}$ and $\numass \leq \qty{0.285}{\eV}$, depending on the \Lya{} implementation we use.
Such data combinations still present an extrapolated minimum at negative values.
They can be put into perspective by considering CMB-only (without lensing) constraints, that stand at \num{0.36} and \qty{0.39}{\eV} for \textit{Planck} PR4 data~\cite{rosenbergCMBPowerSpectra2022,tristramCosmologicalParametersDerived2024}.

Finally, we applied the same formalism to compute profile likelihoods of $m_\mathrm{l}$, the lightest neutrino mass.
This computation becomes ordering dependent, although both orderings yield similar results.
In particular, for DESI DR2 BAO and \textit{Planck} PR4 in \LCDM{}, we find limits at \num{20} and \qty{19}{\meV} in the normal and inverted ordering.

Overall, the minima of the profiles most often lie below zero. 
While \wowaCDM{} constitutes an interesting exception to this trend, we note that the tensions with  physical mass ($\numass \geq 0$) remain very limited in the \LCDM{} cosmological model, rarely exceeding $\chi_0^2 = 1$.
This highlights the need for continued research in this area.
Future neutrino mass analyses will especially benefit from upcoming DESI data, including DR1 \Lya{},  DR2 full-shape and DR3 BAO.

\section{Data Availability}

The DESI data used in this analysis will be made public along the Data Release 2 (details in \url{https:// data.desi.lbl.gov/doc/releases/}).
The data needed to reproduce all figures in this paper are available on Zenodo: \url{https://doi.org/10.5281/zenodo.15878410}.

\acknowledgments

DC, CY and EA acknowledge support from the ANR grant ANR-22-CE92-0037.
MW acknowledges support by the project AIM@LMU funded by the German Federal Ministry of Education and Research (BMBF) under the grant number 16DHBKI013. 
MW acknowledges support from the Excellence Cluster ORIGINS which is funded by the Deutsche Forschungsgemeinschaft (DFG, German Research Foundation) under Germany’s Excellence Strategy – EXC-2094 – 390783311.
JR acknowledges funding from US Department of Energy grant DE-SC0016021.

This material is based upon work supported by the U.S. Department of Energy (DOE), Office of Science, Office of High-Energy Physics, under Contract No. DE–AC02–05CH11231, and by the National Energy Research Scientific Computing Center, a DOE Office of Science User Facility under the same contract. Additional support for DESI was provided by the U.S. National Science Foundation (NSF), Division of Astronomical Sciences under Contract No. AST-0950945 to the NSF’s National Optical-Infrared Astronomy Research Laboratory; the Science and Technology Facilities Council of the United Kingdom; the Gordon and Betty Moore Foundation; the Heising-Simons Foundation; the French Alternative Energies and Atomic Energy Commission (CEA); the National Council of Humanities, Science and Technology of Mexico (CONAHCYT); the Ministry of Science, Innovation and Universities of Spain (MICIU/AEI/10.13039/501100011033), and by the DESI Member Institutions: \url{https://www.desi.lbl.gov/collaborating-institutions}. Any opinions, findings, and conclusions or recommendations expressed in this material are those of the author(s) and do not necessarily reflect the views of the U. S. National Science Foundation, the U. S. Department of Energy, or any of the listed funding agencies.

The authors are honored to be permitted to conduct scientific research on I'oligam Du'ag (Kitt Peak), a mountain with particular significance to the Tohono O’odham Nation.

\bibliography{bibliography}{}
\bibliographystyle{JHEP}

\end{document}